\documentclass[11pt]{article}
\usepackage[utf8]{inputenc}
\usepackage{jheppub}
\usepackage{amsmath}
\usepackage{physics}
\usepackage{tikz}
\usepackage{cleveref}
\usepackage{comment}
\usepackage{natbib}

\usepackage{tikz}
\usetikzlibrary{decorations.pathmorphing}
\usepackage{amsfonts}

\usepackage{subfigure} 

\usepackage{graphics,epsfig}
\usepackage{graphicx}

\usepackage{amsmath}
\usepackage{amssymb}
\usepackage{bbm}
\usepackage{subfigure}
\usepackage{epsfig}
\usepackage{yfonts}
\newcommand{\bea}{\begin{eqnarray}}
\newcommand{\eea}{\end{eqnarray}}
\newcommand{\bean}{\begin{eqnarray*}}
\newcommand{\eean}{\end{eqnarray*}}

\newcommand{\nn}{\nonumber}

\renewcommand{\d}{\delta}

\renewcommand{\d}{\partial}
\newcommand{\be}{\begin{eqnarray}}
\newcommand{\ee}{\end{eqnarray}}

\def\braket#1{\left\langle #1 \right\rangle}

\def\braket#1{\left\langle #1 \right\rangle}

\def\Tr{\mathop{\rm Tr}}

\def\d{{\rm d}}

\def\d{\partial}

\title{Holographic Ordering and Negative entropy in Non-equilibrium Euclidean Black Hole Path Integralsl}
\author{Yang An}
\affiliation{Kavli Institute for Theoretical Sciences (KITS), University of Chinese Academy of Science, 100190 Beijing, P.R. China}
\emailAdd{anyangpeacefulocean@zju.edu.cn}
\abstract{

The Gibbons-Hawking-York (GHY) approach was developed for a Euclidean path integral derivation of equilibrial black hole entropy. To extend it to a near-equilibrium  Euclidean path–integral, we study a static Euclidean shell model.
We calculate the Euclidean action shift for the static simple model thin shell held just outside the horizon, and find agreement with Casini's version of Bekenstein bound.
 We find a negative entropy deficit associated to the gravitational attraction towards the shell. For a holographic interpretation, the deficit corresponds precisely to the apparent horizon area deviation from the extremal surfaces

Therefore, we develop a Euclidean path integral framework in which gravitational force emerges from negative entropy gradients due to Hawking temperature gradients. This setup allows us to introduce Onsager reciprocity and a linear-response relation to build a dissipating system, and treat the configuration as a near-equilibrium steady state (NESS).  This clarify that the gravitational potential is a phenomenon informational and ordering, rather than entropic and disordering.

}

\begin{document}
\maketitle
\section{Introduction}

Black hole thermodynamics provide a profound connection between microscopic degrees of freedom and macroscopic gravitational phenomena. In particular, the Bekenstein-Hawking formula \cite{Bekenstein:1973ur, Hawking:1976de} $S_\text{BH}=A/4G$ 
(with $A$ the horizon area) suggests that black hole entropy counts the hidden quantum microstates of a black hole, as supported by string theory and holographic arguments. 
% Gibbons-Hawking-York approach was used to count such amount of entropy in Euclidean Path integral.

%A general formulation of the Bekenstein bound of quantum field degree of freedom \cite{Marolf:2003sq, Marolf:2004et} , holding for any relativistic quantum field system, was proven in \cite{Casini:2008cr} of Bekenstein Bound that holds for any relativistic quantum field theory, using the non-negativity of relative entropy.

%The Page curve is also an entropic gradient

%Recently, the core of Entropic Gravity is under evolution from being macroscopic and thermal into emergent from quantum field theory degree of freedom. 

In recent years, the idea of emergent gravity has evolved from a purely thermodynamic, macroscopic picture to one rooted in quantum information.
In the framework of AdS/CFT \cite{Maldacena:1997re, Witten:1998qj}, the re-derivation of Einstein equation through holographic entanglement entropy \cite{Ryu:2006bv, Hubeny:2007xt} was carried, to linearized level \cite{Lashkari:2013koa, Faulkner:2013ica}, as well as to non-linear level \cite{Faulkner:2017tkh}, and Jacobson got a new derivation of Einstein equation based on Maximal Vacuum Entanglement Hypothesis \cite{Jacobson:2015hqa}. 
%Jacobson derived semi-classical Einstein Equation in \cite{Jacobson:2015hqa} from entanglement first law and AdS/CFT.
%Entanglement Entropy. , %based on Holographic Principal for ground states, 
While, \cite{Carroll:2016lku} examined the major assumptions in \cite{Jacobson:1995ab, Jacobson:2015hqa} and advances raise a fundamental question: 
what kind of entropy underlies gravity -- is it the thermal entropy associated with horizons, or the entanglement entropy of quantum fields? 
Existing theories often take an entropy extremum or saturation (maximum entropy) condition for granted, implying that gravity arises as a drive toward equilibrium and disordering. 
One major fact against regarding gravitational attraction as such entropic force is that in an idealized free-fall or closed-system process (unitary evolution), the fine-grained entropy need not increase at all; gravitational entropy increase is only observed when we consider dissipation from the perspective of an open system (such as a black hole exchanging heat with a bath).
%what kinds of entropy is involved for gravity: thermal entropy or entanglement entropy? or coarse-graining 

%The answer could probably be the holographic entanglement entropy, which origins from the quantum field theory degree of freedom. 
%We note that even earlier, Casini proved a general version of Bekenstein Bound \cite{Marolf:2003sq, Marolf:2004et} that holds for any relativistic quantum field theory \cite{Casini:2008cr} from the non-negativity of relative entropy.

%These 

In this work, we depart from the traditional ``entropy maximization'' paradigm that gravity is driven by the tendency to maximize entropy (i.e. to increase disorder). 
We propose instead that gravity -- more precisely the gravitational attraction is driven by \textit{negentropy} (negative entropy) -- essentially, the presence of order or information, rather than by the tendency toward disorder. In other words, gravitational potential is treated as a kind of ``negative entropy potential'', where it is ordering (negentropy) that curves spacetime instead of entropy increasing. 
This represents a conceptual shift: rather than gravity being an entropic force pushing a system to maximize entropy, it emerges from an entropy deficit, which we interpret as the negative entropy (negentropy)  that matter introduces relative to the maximum allowed by the Bekenstein bound. In general, one define negative entropy as
\bea
N\equiv -(S_\text{max}-S)
\eea
to characterize the ordering level of the system. $N$ is non-positive since the entropy $S$ below allowed maximum, containing the accessible information or order that in principle can be converted to work.

To formalize this idea, we embed our reasoning in the framework of non-equilibrium thermodynamics, which has been explored to study black hole phase transition \cite{Li:2022oup}. 
In reality, if we hold a mass (``shell'') at some radius outside a black hole, we must continuously perform force against gravity. Maintaining the shell in a fixed, static position is forming a near-equilibrium steady state (NESS). 
The equilibrium viewpoint is incomplete and unnatural for an inherently evolving or steady-state scenario.
In this situation, we remind that Onsager's reciprocity implies a linear-response relationship between thermodynamic fluxes and forces. By placing our thought experiment (a black hole plus an external shell held in place) in a NESS framework, we can identify the gravitational force to emerge from the balance with an entropy-driven force arising from a negentropy gradient, which origins from the temperature gradient, with the dissipative external force for an open system.

%specifically Onsager's theory of linear irreversible processes. This allows us to reconcile the concept of ``gravity as an entropic force'' with a proper treatment of entropy currents and thermodynamic forces. 
%%Previous entropic gravity arguments often invoked a static entropy extremum or saturation condition (assuming an equilibrium state with no net entropy flow) to deduce $F_{\rm grav} \sim T\,\nabla N$ for the gravitational force. 
%
%this system is out of equilibrium and will produce entropy in the environment (for example, as Hawking radiation or heat in a support structure).
%
%with a steady entropy current.

%Maintaining the shell out of equilibrium implies a steady entropy current; Onsager reciprocity gives a linear-response relation
%
%
%By placing our construction into a non-equilibrium steady-state (NESS) framework, we identified the gravitational force as an entropic force arising from the negentropy gradient. 
%
%
%Within the Onsager reciprocal structure, the force-law takes the explicit form:

\subsection{Main results}
In this paper, we address these questions by thoroughly examining perturbations of thin-shell states outside black hole horizons. 
We propose a negative entropic mechanism to describe the gravitational attraction towards such shell of ADM mass $m$, is negative entropy (negentropy). 
\[
F_{\text{grav}} = T_H \,\nabla N,
\] connecting perturbations in the horizon area directly
where $N=-S(M,m)$,
 can reproduce the local inertial force, which by-effectively updates the generalize mechanism of Verlinde's entropic force conjecture \cite{Verlinde:2010hp, Verlinde:2016toy}. 

To make this precise, we exploit the Euclidean path integral picture, by carefully computing the change in the on-shell Euclidean action,
\bea
I_E^{\text{(BH+shell)}} = \frac{\beta_H M}{2} + I_{\text{shell}}~.
\eea
If we introduce a thin shell of matter (or an external reservoir) at a fixed radius outside the horizon, the Euclidean geometry is modified: the presence of the shell shifts the location (and area) of the apparent horizon.  One finds a corresponding change in the horizon area $\Delta A$, acts like an entropy deficit of the Casini-Bekenstein bound
\bea 
N=-\Delta S=-\Delta A(\mu_a)/(4G)\,,
\eea
which we claim essentially origins from the conical defect variation, because the gluing of spacetime by the shell changes defect angle by $\Delta\vartheta$, just like non-equilibrium black hole physics \cite{Li:2022oup}. 
Such negative entropic gradient $\nabla N$ emerges due to the locally isoenergic process in our former work \cite{An:2020ncr}. 

%that perturbing such amount of entropy which can be use to unify the falling tendency. 
%We also observe that the decreasing direction of the Casini-Bekenstein bound $\delta Q=T\delta \Delta S<0$ is the same as the falling tendency during quasi-static. 

%within the Euclidean path integral framework. 

Therefore we suggest that the amount of entropy associated to the gravitational attraction to be
\bea
N=-S(M,m_0 e^{\phi})=-{m_0 e^{\phi} \over T_H}=-8\pi G M m_0 e^{\phi}
\eea
which allows us to adopt a related but distinct viewpoint: we reveal the gravitational potential as a kind of ``negative entropy'' potential, such that gradients in entropy underlie variation of gravitational attraction. Besides, a better holographic meaning can be made to interpret this amount of entropic variation, inspired from \cite{Chandra:2022fwi}, as the area variation of the apparent horizon, to the maximal of coarse-grained entropy.  Remind that here we are working on maximal entangled state of perturbing TFD state \cite{Maldacena:2001kr} instead, we argue the deviation of subdominant $\mu_a$ from extremal surface $\gamma$ is also varied in such amount, which is viewed in AdS/CFT. 
%we would also admit $F_g$ may not be entropic until we get it balanced with $F_{ex}$.

%Verify the bound and area difference.

An intriguing calculation shows adding many infinitesimal static shells and integrating recovers the Bekenstein-Hawking formula
%law in relation of energy $M$:
\[
\int_{0}^{M}\,\frac{dm'}{T_H(m')}=\frac{A}{4G}=S_\text{BH}(M)\,,
%\tag{I1} 
\]
with equivalent horizon area $A$ of a black hole with total mass $M$. This simple calculation shows that the coarse-grained entropy contribution of each shell combines consistently into the full horizon entropy, confirming the internal consistency of the negentropic framework for gravitational attraction.

\paragraph{Structure}
 The remainder of the paper is structured as follows:

\begin{itemize}
\item  In section \ref{Shell}, we set up Euclidean path integrals with thin shells in asymptotic flat Schwarzschild background. We construct the shell geometries by matching regions of two metrics (with masses $M$ and $M+m$) and compute the Euclidean action from the Israel's junction. Then we explain how negative entropy comes out in this setup.

\item  In section \ref{Close-Eq}, we identifies the gravitational force with the entropic gradient. Working in a fixed-shell setup, we derive thermodynamic relations such as $\Delta S = m/T_H$ and $S(M,m)=\Delta A/(4G)\approx 32 \pi G M m $. At last, we elucidate the critical role of external dissipative forces in determining gravitational dynamics, guided by Onsager's reciprocity relations.

% from both the Lorentzian first law and the Euclidean path integral. We show these results agree and interpret them as evidence that mass feels an entropy gradient.
%\item In section \ref{}, 
\item In section \ref{sec:negentropy}, we emphasize how negative entropy changes (work extraction) naturally arise in a coarse-grained description. By grounding gravity in coarse-grained entropy and information theory, our work illuminates the microscopic logic of gravitational attraction and sets the stage for a fully thermodynamic account of spacetime, viewed in AdS/CFT.\end{itemize}
During the context, we take $c=\hbar=k_B=1$ units.

%We relate the redshifted Hawking temperature on the shell to the boundary CFT temperature, and use the modular Hamiltonian formalism to argue that the entropy gradient in the bulk maps to a boundary modular energy flow. This provides a holographic description of our mechanism.

%Free-falling tendency and 2nd law tendency are all natural direction that can't be reversed.

\section{Euclidean Path Integral}
\label{Shell}

We begin by analyzing a black hole with a thin shell of matter using the Euclidean path integral formulation of quantum gravity. This approach will allow us to compute the thermodynamic properties (action, free energy, entropy) of the system in a controlled way, and to identify the entropy deficit caused by the shell. 

In the Euclidean path integral formalism, one considers a Wick rotation $t \to -i\tau$ of the time coordinate, so that the black hole solution is analytically continued to a Euclidean manifold. For a static black hole (such as Schwarzschild), the Euclidean time $\tau$ is compact: one must identify $\tau \sim \tau+\beta$ with period $\beta$ to avoid a conical singularity at the horizon. Physically, this periodicity $\beta$ corresponds to the inverse Hawking temperature $T_H$ of the black hole. In fact, requiring the Euclidean geometry to be regular at the horizon fixes $\beta$ and thus determines $T_H = 1/\beta$. 

For our example, a Schwarzschild black hole of mass $M$ has Euclidean time period $\beta=8\pi G M/c^3$ (later, we use $c=1$ units for simplicity), yielding the Hawking temperature $T_H = 1/(8\pi G M)$. And an external Euclidean evolution will be added due to a shell matter, which shifts the Hawking temperature, and bends the cigar.

%However, when we consider static shell, it is coincident in both Lorentz and Euclidean configuration.

\subsection{Starting from disk: thermal cycle and cigar}

The Euclidean continuation of an eternal black hole results in a smooth cigar-like geometry: in $(r,\tau)$ coordinates, the Euclidean time $\tau$ is a circle that shrinks to zero size at the horizon, much like the tip of a cigar. 
Topologically, the Euclidean black hole spacetime is a disk (a 2D plane with a bounded tip) times the 2-sphere of angular directions, $\Sigma\times S^1_\beta$. At spatial infinity, the Euclidean time circle has circumference $\beta$, matching the periodicity required for temperature $T_H$. This thermal circle $\tau \sim \tau+\beta$ ensures that quantum fields in this background are in thermal equilibrium at temperature $T_H=1/\beta$.

In quantum field theory, a path integral on a Euclidean time circle of length $\beta$ computes the thermal partition function $Z(\beta) = \mathrm{Tr} e^{-\beta H}$, where $H$ is the Hamiltonian. Similarly, in quantum gravity the Euclidean path integral with a periodic time identifies $\tau=0$ and $\tau=\beta$ and corresponds to tracing over all states. Schematically, one can insert a complete set of states at $\tau=0$ and $\tau=\beta$ and sum over them, which yields $Z = \mathrm{Tr}(e^{-\beta H})$. In the gravitational context, $Z$ is dominated by the classical Euclidean geometry (the black hole cigar), so to leading order $Z \approx \exp(-I_E)$ where $I_E$ is the Euclidean action of the on-shell solution. The Euclidean action can be computed by subtracting the action of a reference background (e.g. flat space) to remove divergences at infinity. 

The so-called ``cigar" geometry replaces the disk partition function obtained from tracing the pacman figure because of the curvature of the black hole. The black hole partition function can be represented diagrammatically as
%\begin{figure}[H]
\be
Z_\text{BH}=\begin{matrix}
 \includegraphics[width=2.5cm]{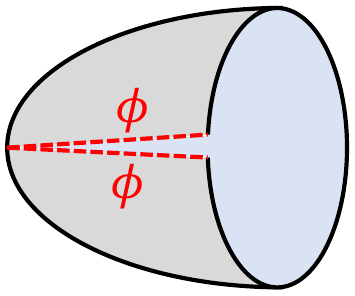}
\end{matrix}=\begin{matrix}
 \includegraphics[width=2.5cm]{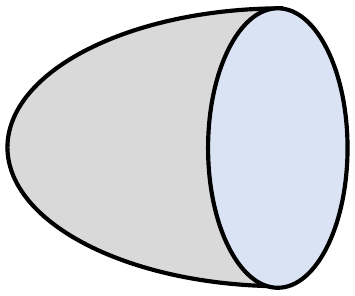}
\end{matrix}\,.
\ee
%\end{figure}\vspace{-1em}

Recall that in ordinary quantum field theory on a spatial manifold $\Sigma$ the Euclidean path integral, representation of a thermal partition function is derived by inserting complete sets of states into a thermal trace
\bea
Z_\text{BH}=\Tr e^{-\beta H}
\eea
After renormalization at large radius $r_0$ with $\mathcal R ^3\times S$ for the flat, which gives $\beta(8\pi r_0-8\pi M)$, the subtraction leaves
\bea
I_E=\beta M/2
\label{SE}
\eea
in agreement with the original derivations by Bekenstein and Hawking. This Euclidean derivation provides a statistical interpretation: the black hole has an enormous entropy proportional to its horizon area, suggesting a huge number of microstates (of order $\exp{A/(4G\hbar)}$) that are not evident in the classical description. In other words, the horizon area acts like a thermodynamic entropy in the gravitational path integral, reinforcing the idea that quantum gravitational degrees of freedom reside on or are associated with the horizon surface.

From the Euclidean point of view, the black hole in thermal equilibrium is described by a Euclidean state often identified with the Hartle-Hawking vacuum. This state can be thought of as the result of evolving the quantum state half a period in Euclidean time and then gluing two copies together.

%\begin{subfigures}
\subsection{A Fixed Shell: bended cigar}

We now introduce a thin spherical shell of matter around the black hole and analyze it using the Euclidean path integral, treating the shell as a perturbation. For simplicity, consider a shell of rest mass $m$ positioned at a fixed Schwarzschild radial coordinate $r_{\mathrm{shell}}>r_s$ (where $r_s=2GM$ is the Schwarzschild radius of the black hole). This shell is held static by an external agent (since a stationary shell outside the horizon is not in free fall and requires support). Physically, this is an open system setup: energy can be exchanged with whatever is holding the shell in place, and the spacetime is not vacuum everywhere. The presence of the shell modifies the metric outside and inside the shell. In fact, the spacetime can be obtained by gluing an exterior Schwarzschild solution of mass $(M+m)$ to an interior Schwarzschild solution of mass $M$ across the shell's worldvolume (using Israel's junction conditions for a thin shell). The result is a metric of the form:
\bea
ds^2 \;=\; \begin{cases}
\Big(1-\frac{2G(M+m)}{r}\Big)d\tau^2 + \Big(1-\frac{2G(M+m)}{r}\Big)^{-1} dr^2 + r^2 d\Omega^2, & (r > r_{\mathrm{shell}}),\\[1ex]
\Big(1-\frac{2GM}{r}\Big)d\tau^2 + \Big(1-\frac{2GM}{r}\Big)^{-1} dr^2 + r^2 d\Omega^2, & (r < r_{\mathrm{shell}}),
\end{cases}
\eea
with appropriate delta-function source at $r=r_{\mathrm{shell}}$ providing the shells' stress-energy. 
In the Euclidean picture, the shell appears as a stationary ring (the red line in the figure above) at radius $r_{\mathrm{shell}}$ extending along the entire Euclidean time circle. Intuitively, one can imagine constructing the Euclidean path integral for this system by starting with the original black hole cigar geometry and inserting the shell's influence: this forces the Euclidean geometry to adjust such that at $r_{\mathrm{shell}}=R$ there is a discontinuity in the derivative of the metric (reflecting the shell's localized energy). One immediate consequence is that the horizon structure changes: with the shell present, there is now an apparent horizon outside the original event horizon which lies at $r_s=2GM$. When a stationary shell of mass $m$ is present at $r_{\mathrm{shell}}$, the spacetime outside the shell is as if there were total mass $(M+m)$ of a black hole.
% If one were ignorant of the shell's support and just looked at the external metric, one would find a horizon at radius $r_{a}=2G(M+m)$ - we can call this the apparent horizon $r_{a}$ caused by the shell's mass. Indeed, outside the shell, outgoing light rays experience the spacetime of a larger black hole and thus have zero expansion at $r=r_a$ . However, the true event horizon $H$ of the spacetime is still at $r_s=2GM$ (inside the shell), because the shell is being held in place and has not collapsed to merge with the black hole. The apparent horizon at $r_a$ lies outside the shell and is a temporary, quasi-horizon from the perspective of an outside observer. If the shell were to be released and fall in, $r_a$ would eventually become the new event horizon after the shell is absorbed. But as long as the shell is static, we have $r_a > r_s$ with $r_a$ acting as the outermost radius where outgoing photons can be momentarily at rest (due to the shell's gravity). This situation is depicted in Figure \ref{bended} (the dashed arc in the figure corresponds to the apparent horizon at $r_a$, which is separate from the true horizon at $r_s$).

\begin{figure}[hbt]
  \includegraphics[width=5cm]{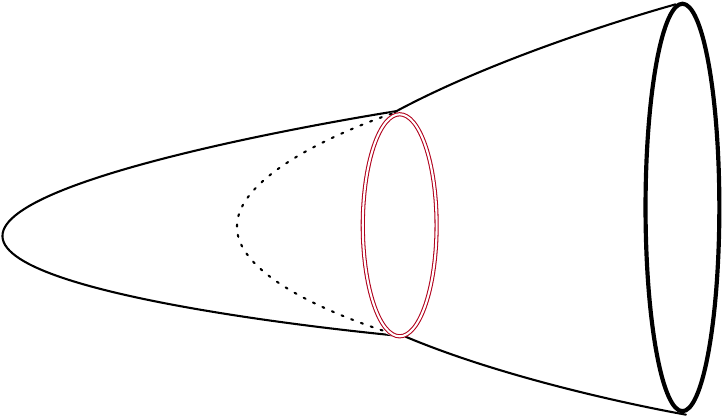}
  \centering
  \caption{The apparent horizon $\mu_a$ at $r_a$ (top of dashed cigar) is apart from $H$ at $r_s$}
  \label{bended}
\end{figure}

%The presence of the static shell does not introduce any conical singularity in the Euclidean geometry as long as the periodicity of $\tau$ is chosen appropriately. We must have the Euclidean time period $\beta$ correspond to the Hawking temperature of the outer metric (mass $M+m$) in order for the geometry outside the shell to be regular at $r_a$. In other words, we set $\beta = 8\pi G (M+m)$ so that the exterior region has no conical defect at $r_a$. This means the system is thermalized at the Hawking temperature $T_H$ of the apparent horizon, which is $T_H = 1/(8\pi G (M+m))$. 

%If the shell is held in place gently (quasi-statically), the entire spacetime can be considered in thermal equilibrium at this temperature. The black hole interior (mass $M$) and the region near the shell (with local redshift factors) adjust consistently so that a single Euclidean period works for the whole configuration. 

%Essentially, the shell radiates or absorbs just enough heat to stay in equilibrium with the black hole's Hawking radiation at that temperature.
%
%What is the entropy of this configuration? 
%Since the shell is stationary and the system is in equilibrium, one might expect that (apart from the shell's own thermodynamic entropy which we take as negligible for a simple mass shell) the total entropy is still given by the horizon entropy. However, which horizon? 
There are two characteristic horizon radii: $r_s$ and $r_a$. The Euclidean path integral approach provides a clue: the dominant contribution to the gravitational action (and thus to the entropy) comes from the geometry at the horizon where the Euclidean time circle closes off. 

In fact, the Euclidean picture in this case is more complicated: the Euclidean time circle has a period $\beta$ that prevents a singularity at $r_a$, but the presence of the shell means that the original horizon at $r_s$ is no longer an accessible Euclidean tip (the shell cuts it off).  Essentially, one finds that to avoid any conical singularity at the black hole horizon, the time period must remain $\beta_H$ (fixed by the regularity at the horizon), but then a conical defect appears at the shell proportional to $m_0$.

\subsection{The Junction Condition}
We now see the inserting of a static thin shell of radius $R$ in the Schwarzschild spacetime.  The interior $r<R$ is itself a vacuum Schwarzschild geometry (of some mass $M$), and the exterior $r>R$ is a Schwarzschild geometry of mass $m$.  The shell at $r=R$ induces a discontinuity in extrinsic curvature. We shouldn't expect the continuum on co-dimensional 1 hypersurface $\Sigma$ of induce metric on both side of the shell, such as
\bea
[h_{ij}]^{\pm}=h_{ij}^{+}-h_{ij}^{-}\neq0.
\label{induce-h}
\eea
with the induced metric
\bea
h^{\pm}_{ij}\,dx^{i}dx^{j}=f_{\pm}(R)\,d\tau^{2}+R^{2}d\Omega^{2}\,,
\eea
with $f_+(r) = 1 - \frac{2G(M+m)}{r}$ and $f_-(r) = 1 - \frac{2GM}{r}$.

 Israel's junction conditions relate this jump to the surface stress-energy $S^i{}_j$ on the shell:
\bea
S^i{}_j = -\frac{1}{8\pi G}\Bigl([K^i{}_j]-\delta^i{}_j[K]\Bigr)\;,
\eea
with $[K] = [K^i{}_i]$ and where $[K^i{}_j]=K^i{}_j{}^+ - K^i{}_j{}^-$ is the difference of extrinsic curvature from outside minus inside.  For a static, spherically symmetric shell this yields a surface energy density $\sigma=-S^\tau{}_\tau$ and an (isotropic) pressure $p=S^\theta{}_\theta=S^\phi{}_\phi$.

 A straightforward calculation yields surface energy density and (isotropic) tension
\bea
\sigma=-\frac{1}{4\pi GR}\left[\sqrt{f_+(R)}-\sqrt{f_-(R)}\right]\,,
\eea
\bea
p=\frac{1}{8\pi G}\left[\frac{f'_+(R)}{2\sqrt{f_+(R)}}-\frac{f'_-(R)}{2\sqrt{f_-(R)}}+\frac{\sqrt{f_+(R)}-\sqrt{f_-(R)}}{R}\right]\,.
\eea

The external agent must supply a radial force $F_{\text{ext}}=4\pi R^{2}p$ to hold the shell. For small perturbation $m\ll M$, we can approximate to get
\bea
\sigma &=&-\frac{1}{4\pi GR}\left[\sqrt{f_+(R)}-\sqrt{f_-(R)}\right]\nn\\
%&= &\frac{\sqrt{f}}{f} G m/R \frac{1}{4\pi GR}+\text O(m^2)\nn\\
&=& \frac{m}{4\pi R^2\sqrt{f(R)}}+\text O(m^2)\,,
\eea
since the rest mass is $m_0\approx \frac{m}{V(r)}$ with the redshift factor $V(r)=\sqrt {f(R)}=\sqrt {f_-(R)}$, one relates
\bea
m_0=4\pi R^2\sigma\,.
\eea
 
 The shell's action can be expressed in terms of its surface energy. One may take the matter Lagrangian density $\mathcal{L}_{\rm shell}=\sigma $ 
 so that 
\bea
I^\text{matter}_{\rm shell} =   \int_\Sigma \mathcal{L}_{\rm shell}\sqrt{h} d^3x \,,
\eea
 where the matter contribution to the action is just
 \bea
 I^\text{matter}_{\text{shell}}= \sqrt{f(R)}\,\beta_H \,4\pi R^{2}\sigma =\beta_H m.
 \label{mshell}
\eea

%We shall check
%\bea
%I_{\text{shell}}^{\text{matter}} = I_{\Sigma}^{\text{net}}\,,
%\eea
%where the net effect $I_{\Sigma}^{\text{net}}$ is calculated from  the GHY terms on the two side of the shell. We leave the detail calculation  in Appendix \ref{ShellAct}.
\subsection{Euclidean Path Integral}

The concept of negative entropy can also be illuminated using the Euclidean path integral formulation of gravity. In the Euclidean approach, thermodynamic properties of black holes (and gravitating systems in general) are derived by analytically continuing to imaginary time $\tau = it$ and demanding periodicity in $\tau$ with period $\beta = 1/T$ to represent a finite temperature $T$. The partition function is formally given by a Euclidean path integral $Z = \int \mathcal{D}g\,e^{-I_E[g]}$, where $I_E$ is the Euclidean gravitational action (including any matter contributions). A stationary point of $I_E$ (a classical Euclidean solution) yields the dominant contribution $Z \approx e^{-I_E^{\text{on-shell}}}$. The thermodynamic free energy $F$ is related to the on-shell action by $I_E = \beta F$, and the entropy can be obtained via $S = \beta (E - F)$ or by the conical singularity method (differentiating with respect to the period $\beta$) .

For the Schwarzschild black hole, the Euclidean section is a smooth manifold only if $\beta = \beta_H \equiv 1/T_H = 8\pi G M$. The Euclidean geometry looks like a ``cigar '': the Euclidean time $\tau$ is periodic with period $\beta_H$, smoothly capping off at the horizon (which in the Euclidean signature is a regular point, not a true singularity). The on-shell Euclidean action for the Schwarzschild solution is $I_E = \frac{\beta_H M}{2}$. From this, one can derive the thermodynamic quantities: $E=M$, $F = M/2$, and $S = \beta_H (E - F) = \beta_H (M - M/2) = \frac{\beta_H M}{2} = \frac{1}{2} \beta_H M$. Inserting $\beta_H = 8\pi G M$, we get $S = 4\pi G M^2$, which indeed equals $A/(4G)$ for the Schwarzschild black hole. This is the standard Euclidean derivation of the Bekenstein-Hawking entropy. We noted here, one observation is
\bea
N=-(\beta_H F-S)=0\,,
\label{N-vanish}
\eea
for later convenience. No one can subtract useful work from such black hole for it is maxima-mixed.

Now, how does negative entropy appear in this Euclidean framework? In the Euclidean path integral, a configuration with a mass $m$ held at some radius outside the black hole corresponds to a Euclidean solution with a ``defect angle'' or localized curvature at the location of the mass. One way to construct such a solution is to start with the Euclidean black hole metric (the cigar geometry) and introduce a thin shell of matter at radius $r$. The presence of the shell will in general change the periodicity required for smoothness of the Euclidean section. If the shell is held in place (for example, by an external pressure or by a strut), the Euclidean geometry can be thought of as a black hole solution for $r<r_{shell}$ matched to a different spacetime (one with a conical defect corresponding to the shell's mass) for $r_{shell}$. This ``gluing'' results in a conical angle defect at the shell proportional to the shell's rest mass $m$ (proper mass). Intuitively, if the Euclidean time period were $\beta_H$ everywhere, the introduction of the shell would create a mismatch -- either the exterior or interior near the shell would not fit perfectly in the $2\pi$ periodic time unless we allow a conical defect.

The relation between the defect angle and the shell's mass can be derived from the Euclidean Einstein equations or equivalently via the Israel junction conditions at the shell.
The Euclidean time period must remain fixed at $\beta_H=8\pi GM$ to avoid any conical singularity at the horizon. 

 Consequently, at the shell radius $R$, a localized conical defect angle $\Delta\vartheta$ arises, proportional directly to the shell's intrinsic mass $m_0$: this defect angle $\Delta\vartheta$ is related to the mass $m_0$ by $\Delta\vartheta = 8\pi G\,m_0$ (this can be derived from the Gauss-Bonnet theorem or the Israel junction condition on the shell in Euclidean space). The Euclidean action acquires an extra contribution due to the shell's presence. In fact, one can show that the on-shell action for the composite configuration is 
\bea
I_E^{\text{(BH+shell)}} = \frac{\beta_H M}{2} + I_{\text{shell}}~,
\eea
where the first term is the pure black hole part and $I_{\text{shell}}$ is an additional contribution coming from the shell's stress-energy. For a thin shell of mass $m$ at radius $R$, from (\ref{mshell}) matter energy contribution we shall suggest that $I_{\text{shell}}$ turns out to be 
\bea
I_{\text{shell}} \approx \beta_H \,{m_0\,e^{\phi(R)}}~,
\eea
where $e^{\phi(R)} = \sqrt{1-\frac{2GM}{R}}$ as before (this factor effectively accounts for the redshift of the shell's energy in the Euclidean geometry). 

From the action, we can extract the free energy and entropy. The total free energy of the black hole + shell system is $F = F_{\text{BH}} + F_{\text{shell}} = M/2 + {m_0\,e^{\phi}}$. The internal energy (ADM energy) is $E = M + m_0\,e^{\phi}$ (since the redshifted mass $m_0\,e^{\phi}$ is the contribution of the shell to the total energy as seen from infinity). Now, using $S = \beta_H (E - F)$, we get 

\bea
S = \beta_H \Big[(M + m_0\,e^{\phi}) - \Big(\frac{M}{2} + {m_0\,e^{\phi}}\Big)\Big] = \beta_H \frac{M}{2} %+ {m_0\,e^{\phi}}\Big)~. 
\eea
We notice though the free energy is lifted by the shell, the entropy remains the same as $S_\text{BH}(M)$ of a black hole of mass $M$.
%\footnote{But for quasi-local test matter field as in \cite{Brown:1992bq,Brown:1992br}, %self-gravitating 
% calculation may still gives
%\bea
%I_m= \beta_H m/2\,.%I_{\text{shell}}^{\text{matter}} + \Delta I^{\text{GHY}}_{\infty}=
%\eea
%for which the entropy is also shifted by quadratic. Our model is in different fixed setup with considering back-reaction of the shell.
%%where $\Delta I^{\text{GHY}}_{\infty}=-\beta_H m/2$ is shifted from a black hole of mass $M$ for the GHY boundary term at infinity.
%} 

Define such deficit term as $N$
\bea
N=-(\beta_H F-S)=-\beta_H m\,,
\eea
without arguing it is negative entropy for now. We find this deficit term $N=-8\pi G M m_0\,e^{\phi}$ is precisely $-\frac{\Delta A}{4G}$ where $\Delta A = 32\pi G^2 M m_0\,e^{\phi}$ is the horizon area increase if the shell fell in. 

For comparison, to (\ref{N-vanish}) we mentioned earlier.
\bea
N=0\,,
\eea
 for a black hole of mass $M$.

\subsection{Apparent Horizon}
 
 The apparent horizon $\mu_a$ is the outermost of all trapped surfaces, also called the marginally outer trapped surface. 
Mathematically, the apparent horizon is defined by the condition that outgoing null congruences have zero expansion  at that surface, which means $\theta_k \equiv \nabla_k \ln(\mathrm{Area}[n]) = 0$, where $k$ is a future-outwards null vector, and $n$ is a small pencil of light rays shot out in
the $k$ direction from a small neighborhood of a point on $\mu_a$.

From outside of the shell, the expansion $\theta_k = \nabla_\mu k^\mu\propto 1-\frac{2G(M+m)}{r}$, we have the apparent horizon $\mu_a$ at 
\bea
r_a=2G(M+m)\,.
\eea It is just as if a event horizon position of a bigger black hole, since the mass shell bends the light.

For small shell mass $m\ll M$, we would define the entropy associated to such attraction as
\bea
S(M,m)\equiv\Delta S= {\Delta A\over{4G}}\approx 8\pi G Mm
\eea
 where the area difference is 
  \bea
 \Delta A=A(\mu_a)-A(H)\approx 32\pi G^2Mm+O(m^2)
 \eea
 with 
 \bea
A(\mu_a)=4\pi r_a^2=16\pi G^2(M+m)^2, \quad A(H) = 4\pi (2GM)^2.
\eea
from the event horizon $H$ as the dominate extremal surface $\gamma$ at $r_s=2GM$.

We just verify to the order of O(m), coarse-graining entropy increase by $$S(M,m)=\Delta A / 4 G=\Delta E/T_H\,,$$ which is just a simple saturation version of the Casini-Bekenstein bound from a totally thermalized maximal energy.

 But the actual entropy of the configuration, as given by the path integral, is not $S_{\text{coarse}}$ if the shell is maintained in a non-equilibrium state. The path integral above effectively assumed the system is in equilibrium at a single temperature $T_H$ (since we enforced the horizon regularity and single $\beta_H$). If the shell is held in place without exchanging heat with the black hole (e.g. an adiabatic boundary condition at the shell), the correct way to account for the shell's degrees of freedom is different: one would treat the shell's microstates separately, or consider a microcanonical ensemble. In such a case, the entropy associated with the shell would be lower than the $\Delta S_{\text{BH}}$ computed for full thermal equilibrium. The difference $S(M,m)=S_{\text{coarse}} - S_{\text{vN}}$ may possibly show up as a consequence of a \emph{negative mode} or instability in the Euclidean action.

 The linear term $8\pi G M m$ is exactly the $\Delta S$ we have identified. Thus, our stationary shell analysis matches the standard Bekenstein-Hawking result to first order and provides a concrete, controlled verification of the entropy accounting for matter just outside a horizon. In summary, the apparent horizon's area increase $\Delta A$ serves as a holographic interpretation for the entropy associated with the shell.%, reinforcing the viewpoint that gravity encodes thermodynamic information at horizons.

 \paragraph{Bekenstein-Hawking Entropy in the attraction}

Viewing each layer differing by a Euclidean time evolution $\delta \beta \propto \delta m$, we find coincidentally the integrating many infinite thin shell will reproduce
\bea
\int_0^M  S(m, dm)= 4\pi G M^2\equiv A/4G=S_\text{BH}(M)\,,
\eea 
the result of accumulation will results into an area equality even it is not a black hole. Such a result is a strong consistency check, showing that our coarse-grained entropy function $S(M,m)$ can be integrated to yield the correct total entropy. 

This simple calculation underscores that our formulation is not just qualitatively but quantitatively in line with known gravitational thermodynamics from micro into macro, promisingly transforming the observable gravitational phenomenon into informational content.

 \section{Near-Equilibrium Situations for Gradient}
 \label{Close-Eq}

 Having established the notion of negative entropy $N$ in a gravitating system, we now transition to a description that treats gravity as an emergent thermodynamic force for \emph{near-equilibrium steady state} (NESS). We will recast our scenario (a black hole with an external shell) in the language of entropy flows and thermodynamic forces. This section builds a bridge between the Euclidean (equilibrium) analysis and a real-time, non-equilibrium perspective.  
 
%In the sense that the connection and continuity of spacetime geometry is closely related to the entanglement structure of QFT states,\cite{Swingle:2009bg,VanRaamsdonk:2009ar,VanRaamsdonk:2010pw} show entanglement builds spacetime geometry when increasing level of entanglement. But as a natural property of von Neumann entropy, this fine-grained quantity is conserved under unitary time evolution. 

In the first half, we review our former special method called ``Perturbing on the Perturbation" technique for evolving excited state in quasi-equilibrium, and reinterpret this development of emergent gravity theory as the evolution of Euclidean states with the bath coupling.  And only in this special case to evolve euclidean state in quasi-static, the entropy associated with gravitational attraction coincide to holographic entanglement entropy bound and coarse-grained entropy with a good holographic meaning.

In the second half, we are going to check entropic force formula
\bea
F_\mu=T\nabla_\mu S\,.
\label{entforce}
\eea 
is an thermodynamic force, which indeed reproduces temperature gradient in Tolman relation. Between different NESS of the shell, the entropy can be generated. And we will translate this formula (\ref{entforce}) in Onsager's linear response framework.

The final goal is to rigorously examine the generation and maintenance of temperature gradients within gravitational systems treated as open thermodynamic environments.

 \subsection{Review: Perturbing Entanglement 1st Law on the Chaotic Vacuum}

Spacetime is emergent dual to the quantum entanglement. On such background, the amount of fine-grained entropy is usually less than the maximal entropy bound.

What really matters here are two following points: 

(i) We develop a modular Hamiltonian approach to extract gravity as an extra work term that we seek from the entanglement first law.
The entanglement first law states that if $\rho_R(\lambda)$ of a state in the subsystem $R$ varying with one parameter $\lambda$, to the first order perturbation $d\lambda$ at $\lambda=\lambda_0$, we always have the following equation
\bea
\frac{dS(\rho_R)}{d\lambda } =\Tr \left(\frac{d\rho_R}{d\lambda} {K_0}\right)
\label{firstlawK}
\eea
or we can rewrite it as
\bea
dS=d\braket{K_0}
\eea
where $K_0=-\log{\rho_R(\lambda=\lambda_0)}$ is the modular Hamiltonian of the initial state. A detail proof can be find in \cite{VanRaamsdonk:2016exw}. As a consequence of (\ref{firstlawK}) between the excited state and vacuum state, we show a local force can be reproduced in an entropic mechanism
\bea
F_{\mu}=T(r)\nabla_\mu \braket{K_0-K_1}_1\,.
\label{Fg}
\eea
as the contribution exactly from the difference between the excited state modular Hamiltonian $K_1$ and the vacuum $K_0$, where $T(r)=\frac{\kappa}{2\pi V(r)}$ represents the local temperature with the surface gravity $\kappa$ and the redshift factor $V(r)$.
We already know the modular Hamiltonian $K_0$ of the Hartle-Hawking state reduced in external of horizon
\bea
\rho_{HH}=\frac{e^{-\beta H}}{Z}
\eea
is simply $K_0=\beta H=H/T_H$, which is exactly the operator for the time Killing vector $\partial_t$, while the excited modular Hamiltonian is generally of the form $K_1=K_0+O$ in \cite{Balakrishnan:2020lbp,Arias:2020qpg} Compare with the thermodynamic first law $dW+dQ=dE$, one can easily make the hypothesis that the work term is related to
\bea
dW_g=-Td \braket {O}_1\,.
\eea

The expression (\ref{Fg}) can reproduce the same covariant expression for the gravitational force $\bold{F}_{g}$ in GR as tested in the spacetime of asymptotic flat Schwarzschild black holes and Newton's 2nd Law $F=ma$ for Rindler space. This mechanism works generally beyond near horizon region, and it also turns the derivation of Einstein equation from null screen \cite{Jacobson:1995ab} to the time-like screen \cite{Verlinde:2010hp} valid, as we later proved in \cite{An:2020ncr}. Using the Entanglement 1st Law, we derived the semi-classical expressions for a thermodynamic force $F_\mu$ (\ref{Fg}). We consider the specific situation when the relative entropy vanishes, which means the saturation of the Casini-Bekenstein bound (\ref{CBbound}) as in \cite{Casini:2008cr}.  
 
Therefore,  the resulting temperature gradient $\nabla T$ for the local observer Alice is the reason for the thermodynamic force $ F_{\mu}$ as well as an entropic gradient $\nabla S$, in the spirit of the Onsager reciprocity relations. Or to the view an asymptotic observer Bob,  the temperature is fixed to be $T_H$ but the ADM energy gets perturbed during the process such that the isoenergic process for Alice becomes isothermal process for Bob.  % for non-equilibrium thermodynamics.%Thus the existence of the external force measured by accelerating observers mountains its frequency not redshifted, when moving to a nearby position quasi-statically in the global causal wedge. 

% will have entropic gradient % to form specific thermodynamic processes. 
%Ever since the laws of black hole thermodynamics was proposed by Hawking \cite{Bardeen:1973gs}, spacetime was regarded to be emergent from microscopic states which has entropy and temperature. Black holes have Bekenstein-Hawking entropy \cite{Bekenstein:1973ur} and Hawking temperature\cite{Hawking:1974sw}. Unruh was the first to understand the Hawking Temperature as a observer-dependent phenomenon \cite{Unruh:1976db} by considering accelerating observers in Minkowski spacetime, where accelerating observers will find the usual Minkowski vacuum thermalized with particles.

(ii) We show the gravitational force is not entropic by keeping entropy and thermal distribution invariant during free-falling, until an isoenergic process happened to overcome the gravitational redshift effect. 
%=e^{\phi(r)}$ to the generalized Newton's potential $\phi$.%=\frac{1}{2}\log{(-\xi^\mu\xi_\mu)}$.
%Based on \cite{Marolf:2004et,Casini:2008cr}, we relate the inertial force to the entropy and energy of the exited states subtracted from thermal vacuum% in free field theory.  

The Casini's version of the Bekenstein \cite{Casini:2008cr} bound is
\bea
 \Delta S\leq \frac{\Delta \braket{H}}{T_H}=\frac{\Delta \braket{H_A}}{T}=\lim_{upper}\Delta S\,,
\label{CBbound}
\eea where $\Delta S=S_1-S_0$ and $\Delta \braket{H_A}=\braket{H_A}_1-\braket{H_A}_0$ for the local $H_A=H/V(r)$ and local temperature $T=T_H/V(r)$ are the renormalized entanglement entropy and the renormalized energy respectively of excited state $\rho^1_R$ from the vacuum $\rho^0_R$. When this bound is saturated, (\ref{Fg}) leads to the thermodynamic expression

\bea
%F_{\mu}(r)=\frac{1}{1\pm e^{-\omega/T}}\frac{\omega}{T}\frac{\partial T}{\partial r} \delta_\mu^r \,,
F_\mu=\lim_{upper}\Delta S\nabla_\mu T= \frac{\nabla_\mu T}{T}\Delta \braket{H}\,,
\label{uniexforce}
\eea
for $F_{\mu}$ to compare with the local inertial force $\textbf{F}_{g}=-ma_\mu$ and the entropic force formula, where the renormalized energy $\Delta \braket{H}$ is statistics-dependent with the statistical factor $\frac{1}{1\pm e^{-\omega/T}}$.
%%\bea depends on whether it is Bose-Einstein or Fermi-Dirac statistic
%%\frac{\omega}{T} \frac{\partial T}{\partial r}\delta_\mu^r=- {G\M\om\over{r^2(1-\frac{2G\M}{r^2})}}\delta_\mu^r\,,
%%\eea
%the expression approximates to the local inertial force when applying local Hawking temperature $T(r)=T_H/V(r)$ to $T$. It helps to explain the origin of entropic gradient proposed, since such local measurement for the Hawking temperature supply a natural temperature field during the specific process, which is related by a redshift factor. 
After adopt $T=\frac{T_{H}}{V(r)}$ with 
\bea
V(r)=e^{\phi(r)}
\eea being the redshift factor to the generalized Newton's potential $\phi$, this expression can reproduce exactly the local gravitational inertial force and Newton's 2nd Law in generic, as shown in appendix.

\subsection{Near-equilibrium steady state for a black hole with shell}

% in QFT. %The existence of event horizon is responsible for this only thermodynamic effect. In this very specific situation, which we don't need any knowledge beyond the static mechanics, can we reveal the gravitational force from thermodynamics for the microscopic particles states thermalized in this temperature field?
%Noticing the Hawking temperature is position-dependent for the thermal situation seen by different local static observers 
The term \emph{near-equilibrium} means the system is close to thermal equilibrium but with a small drive that causes a steady flux. In our case, imagine the shell is held at radius $R$ by an external string, and the system is allowed to exchange heat with a distant bath so that it settles into a steady state. If the shell is not at absolute rest but maybe slowly lowering, one can have the quasi-static process. 

%From above, we see a string exerting external force is necessary role to  hanging such thin shell as test particle quasi-statically. From the view of thermodynamics of black hoe system, we just make a proper Near-Equilibrium situation. 
\begin{figure}[h]
		\centering
  \includegraphics[width=4.5in]{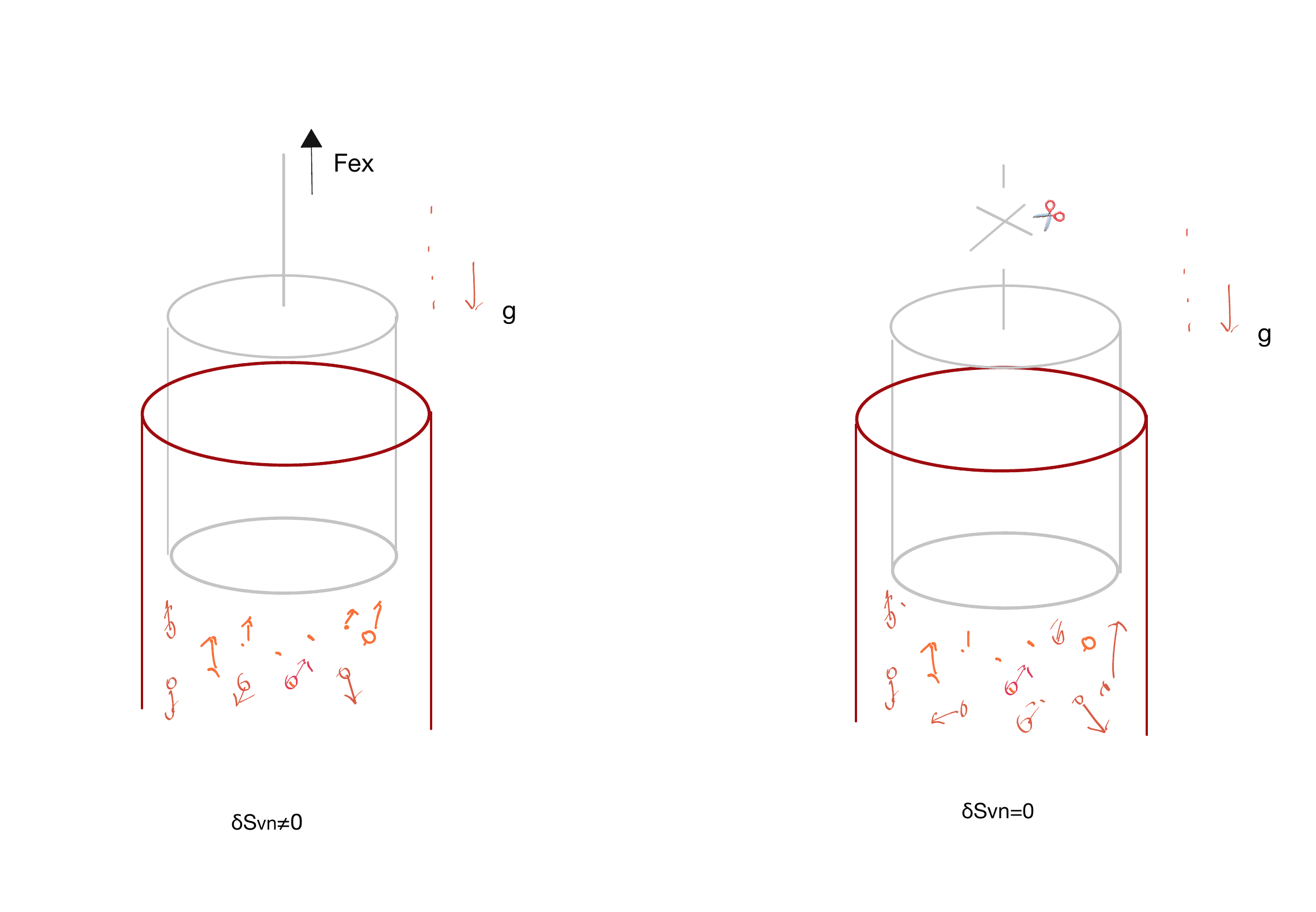}
  \caption{
  %When the gravitational potential completely transform into thermal energy, Whether there is an external force, will $\delta S=0$ or not? 
 In an isolated system without external force by the top string (Right), gravitational evolving such as free falling doesn't change fine-grained entropy even some of gravitational potential gets dissipated thermally. While, entanglement entropy can be changed with transforming energy into external system (Left). 
 One way to tell falling tendency through entropy variation, is to bring in the presence of external force for equilibrating, and then take the split from the external system. 
 %and its direction shows the opposite direction of gravitational force.
  }
  \label{engines}
\end{figure}

To make an analogy to a thermodynamic situation, we can show this gravitational falling situation as a reversible ideal heat engine in Figure \ref{engines}. 
With the equilibrating by external system, one could evolve Euclidean states in quasi-static processes like reversible heat machine. Then, we will orient the variation of the entropy bound in the direction of increasing heat flux.

The perturbing of test particle to the nearby non-inertial trajectory in the global causal wedge can be done by cancelling the  gravitational redshift effect. This is equivalent to mountain the frequency $\omega$ fixed for the single-mode oscillator model. 
At the same time, the existence of the external force balanced with the gravitational force to $\bold{F}_\text{ext}+\bold{F}_\text{grav}=0$, could result in a local temperature gradient $\nabla T$ measured by accelerating observers.

Under the semi-static process to extract gravitational force by fixing local measurement of $\Delta \braket H\rightarrow m$ for nearby static observers,
we show the saturation of the entropy bound 
%as an alternative holographic condition. 

\bea
\Delta S= {\Delta \braket{H}\over T}\,,
\label{BCbound}
\eea
leads to an entropic gradient generally
\bea
\nabla_\mu S= \frac{m}{T_0}\nabla_\mu V\,,
\label{entgrad}
\eea
where $m$ is the mass of the test particle and $V=e^\phi$ is the redshift factor with respect to the general gravitational potential $\phi$,
% depending on the position, 
while $T$ standing alternatively for the local measurement of the Unruh temperature or the Hawking temperature for static observers along with $T=T_0$ for $V=1$. It proves the necessity of external force for an entropic mechanism.
%show under this condition

By defining the generalized gravitational potential
$\phi\equiv \frac{1}{2}\log\{-\chi_\mu\chi^\mu\}$ for time killing vector $\chi^\mu$ and writing redshift factor as 
$V=e^\phi=\sqrt{-g_{tt}}$, we can rewrite our results (\ref{entgrad}) as
\bea
\nabla_\mu S&=&\frac{1}{T_H}me^\phi\nabla_\mu\phi\,.
\label{nSphi}
\eea
with local temperature
\bea
T&=&T_He^{-\phi}\,.
\label{Tphi}
\eea
To explore an entanglement story for gravitational force through detailed calculation, our early work \cite{An:2018hyt} brought in the toy model in \cite{Marolf:2003sq, Marolf:2004et, Casini:2008cr}, the single-mode entangled oscillator reduced in the subsystem, to consider when the black hole background geometry thermalizes the oscillators in the local Hawking temperature, in replacement of the  common role of the Unruh temperature in common local approaches of Entropic Gravity theories such as in \cite{Jacobson:1995ab, Verlinde:2010hp}.

\subsection{Distinguish Coarse-graining direction of Adiabatic from Quasi-static} 

%In thermodynamics, one way to tell the evolving direction of adiabatic process is to switch the system quasi-static.
%The second law says entropy in an \emph{isolated} (adiabatic) system never decreases. But  for local subsystems without an external reference,, the direction of time or entropy tendency can be ambiguous to tell. By introducing a quasi-static split with an external system, the NESS is also obtained, and we can observe  {entropy production explicitly}.

In classical thermodynamics, the arrow of an adiabatic evolution is fixed by the second law: the total entropy of an isolated system cannot decrease.  Yet for a local subsystem--devoid of any external reference-- the sign of the entropy change, and hence the ``direction of time'', can be ambiguous. A standard way to resolve this ambiguity is to place the subsystem in gentle contact with a large reservoir and drive it through an almost-reversible (quasi-static) deformation. 
\begin{figure}[hbt]
		\centering
  \includegraphics[width=3.4in]{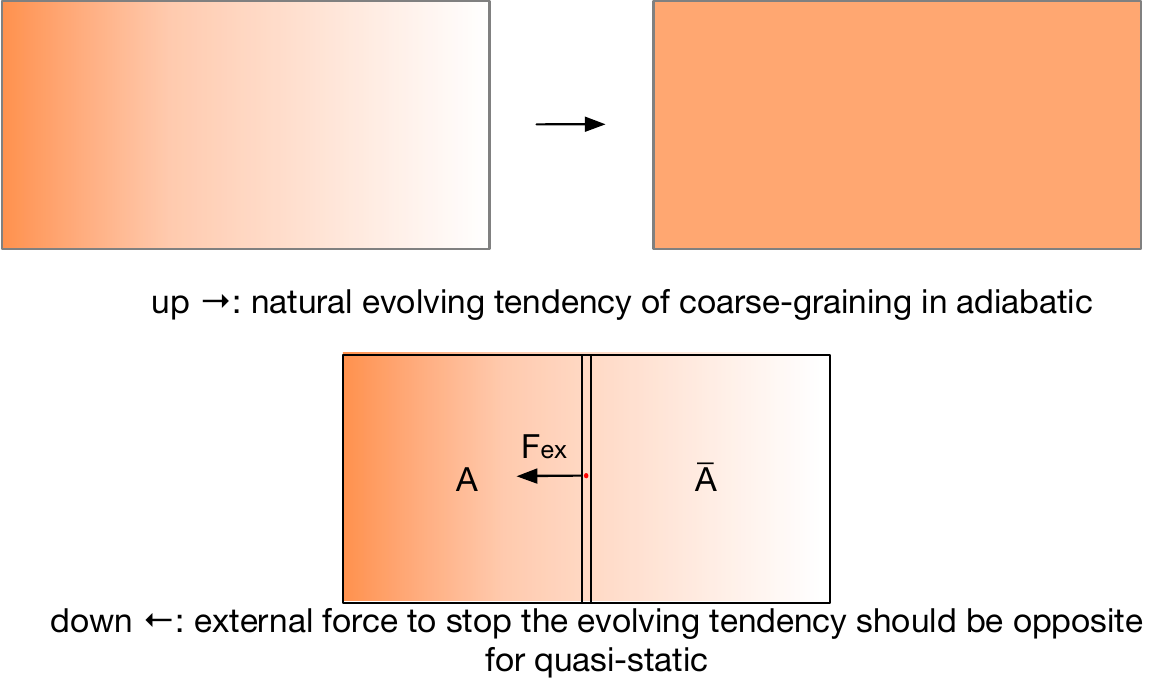}
  \caption{
  From the view of Onsager reciprocity relations for dissipating processes out of equilibrium, the density gradient/temperature gradient gives thermodynamic force, which can be stop by an external force. After we equilibrate the isolated system bipartitely with $F_{ex}$ to stop the evolving, $F_{ex}$ ($\leftarrow $) is always in the opposite direction of the evolving direction ($\rightarrow $). In the same way to gravity, if we stop the free-falling for quasi-static, we are able to tell the falling tendency from opposite direction of  $\bold F_{ex}$.}
  \label{Evolve}
\end{figure}

At some time before equilibrium, one can freeze the evolution of system by adding a piston with external force $F_{ex}$ for equilibrating and split the whole system bipartitely into $A$ and $\bar A$, so one have
\bea
F_{{A}-\bar{A}}+F_{ex}=0
\label{balance}
\eea
where $F_{{A}-\bar{A}}$ in the middle is emergent from such as the density gradient between two subsystems. 

The original tendency could be distinguished as the opposite direction of the external force, as we show in Figure \ref{Evolve}. After the split, $\bar A$ plays the same role of the bath, which includes all the environment as well as  $F_{ex}$ in $\bar A$ to evolve subsystem $A$.
% One gets thermodynamic 1st law to subsystem $A$
%\bea
%dW+ TdS=d E\,.
%\eea

The external agent then plays two roles simultaneously: firstly,	it supplies (or extracts) just enough work or heat to keep the composite system in a near-equilibrium steady state (NESS);
Secondly, it provides a bookkeeping reference against which the entropy production of the subsystem can be measured unambiguously.

With the split in place, every infinitesimal step produces a well-defined, positive entropy increment in the combined ``subsystem + bath'', making the irreversible character of the process manifest.

\paragraph{Analogy to gravitational falling}

From above, we see string exerting external force is necessary role to  hanging such thin shell as test particle quasi-statically. From the view of thermodynamics of black hoe system, we just make a proper near-equilibrium situation.

Halting the free-fall by inserting a ``piston'' (or applying an external force) freezes the configuration, creating two subsystems. At that moment, an imbalance in intensive parameters (temperature, pressure) reveals the system's natural direction of evolution. 
%For the entropy bound is saturated, as in the static case,
 %, where the coarse-graining takes place to maximize the entropy. %$\delta S_\text{bath}>0$.
  And one can use direction of $F_{ex}$ shows this falling tendency. Switching back to the adiabatic process, once we release the piston freely. As Figure \ref{engines} shows.

\begin{figure}[hbt]
		\centering
  \includegraphics[width=5in]{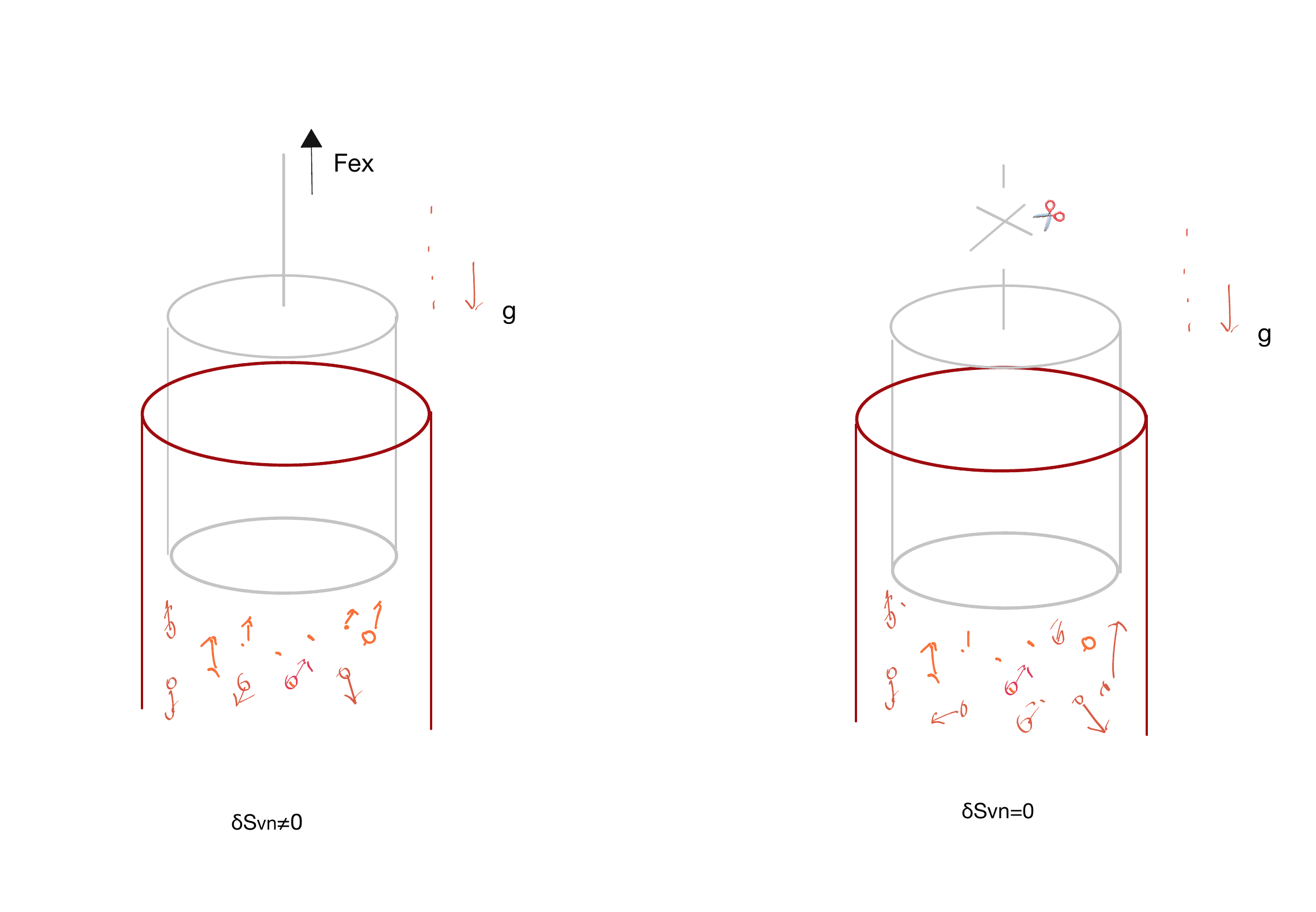}
  \caption{
  %When the gravitational potential completely transform into thermal energy, Whether there is an external force, will $\delta S=0$ or not? 
 In an isolated system without external force by the top string (Right), gravitational evolving such as free falling doesn't change fine-grained entropy even some of gravitational potential gets dissipated thermally. While, entanglement entropy can be changed with transforming energy into external system (Left). 
 One way to tell falling tendency through entropy variation, is to bring in the presence of external force for equilibrating, and then take the split from the external system. 
 %and its direction shows the opposite direction of gravitational force.
  }
  \label{engines}
\end{figure}

The perturbing of test particle to the nearby non-inertial trajectory in the global causal wedge by cancelling the  gravitational redshift effect, is equivalent to mountain the frequency $\omega$ fixed for the single-mode oscillator model. 
At the same time, the existence of the external force balanced with the gravitational force to $\bold{F}_{ex}+\bold{F}_{g}=0$, could result in a local temperature gradient $\nabla T$ measured by accelerating observers.

\subsection{Introducing Onsager Reciprocity to Non-Equilibrium Gravity}

Let $J_i$ denote the fluxes (with $i=1,2$ for the thermal and mechanical channels, respectively) and $X_j$ the corresponding thermodynamic forces. 
Near equilibrium, the linear--response ansatz reads
\bea
 J_i = \sum_j L_{ij} X_j 
\eea
with $L_{ij}$ being the Onsager response matrix. By Onsager reciprocity which is also called Onsager-Casimir symmetry, the cross-coefficients are symmetric, $L_{12}=L_{21}$, since our system is suggested time-reversal invariant at microscopic level (like common reversible system without magnetic field or Coriolis forces breaking reciprocity).

\paragraph{Thermodynamic forces and currents}

To quantify the non-equilibrium dynamics, we introduce two pairs of thermodynamic fluxes and forces for the static thin-shell system. The first pair is the heat (entropy) flow versus temperature gradient, and the second is the shell's radial motion versus the gravitational (generalized) force. In the near-equilibrium steady state, these currents and forces satisfy Onsager's linear response relations. 

We begin by defining two macroscopic force that arise naturally
in a static, spherically symmetric metric of the form $ds^{2}=-e^{2\phi(r)}dt^{2}+e^{-2\phi(r)}dr^{2}+r^{2}d\Omega^{2}$.%Since the local temperature is $T(r) = T_H/\sqrt{-g_{tt}(r)}$:

%In non-equilibrium thermodynamics, a central role is played by the notion of thermodynamic forces, fluxes, and their induced entropy production. Within the linear response regime, the entropy production is expressed as the sum of products between thermodynamic fluxes $J_i$ and conjugate thermodynamic forces $X_i$:
%\bea
%\sigma \;=\; \sum_i J_i X_i \;\geq\; 0\,.
%\eea

\paragraph{Thermal pair:} 
In classical thermodynamic contexts, such as heat conduction or thermal convection, the local thermodynamic force for heat conduction is usually identified as the spatial gradient of inverse temperature:
\bea
X_{T} \;=\; \nabla\left(\frac{1}{T}\right)\,,
\eea
where the force pointing towards lower temperature.%, paired with the heat flux $J_{S} = -\lambda\nabla T$, where $\lambda$ is thermal conductivity. The negative entropy $N$ is transformed into $\delta S$ external bath by dissipating. 

The flux $J_1$ conjugate to $X_T$ is the entropy current (or heat flow) $J_S$, representing the outward flow of entropy (or equivalently, inward flow of negentropy) in response to a temperature difference. 
When moving the shell with external force towards horizon, such steady state can be produced with heat flux.
 Physically, this corresponds to energy dissipation that increases the entropy of the external bath. The negative entropy $N$ is transformed into $\delta S$ external bath by dissipating.

\paragraph{Mechanical pair:} The force $X_G$ is the force associated with the shell's position, essentially the gravitational pull on the shell. We can write
\bea
X_G \equiv \frac{\nabla\phi}{T}\,,
\eea
for generalized force $X_2$ with equal unit as $X_1$, and we have the following equality
\bea
m T X_G= F_{\rm grav} = \frac{dW_g}{dr}\,,
\eea
where $W_g$ is the gravitational potential (work) energy of the shell. For example, if $m$ is the shell's  mass, $\bold F_{\rm grav} \approx m\nabla\phi(r)$ for a shell at radius $r$. The associated flux $J_2$ is the radial velocity of the shell $v = dr/dt$, which measures the shell's motion ( $v<0$ inwards or $v>0$ outwards). In the strictly static state $v=0$, but we consider small deviations (quasi-static drift or oscillation of the shell's position) as the mechanical current. Intuitively, $X_G$ is like a pressure difference or force trying to move the shell, and $J_2$ is the response (shell's velocity or displacement rate).

\paragraph{Balance condition}

 In a static configuration, $X_G$ is balanced by the external agent's force $F_{\rm ex}$ holding the shell up.

Being NESS moving with small deviation of velocity $v$, we will show the balance means that the gravitational field itself naturally generates a spatial temperature gradient through the Tolman-Ehrenfest relation, which states that a static gravitational potential induces equilibrium temperature variations:
\begin{equation}
X_{G}\;=\; -\nabla\Bigl(\tfrac{1}{T}\Bigr),
\qquad
%X_{\phi}\;\equiv\; \frac{\nabla\phi}{T}, \label{eq:forcesDef}  more precisely the gradient of inverse temperature $\beta$
\end{equation}
where in our case $T(r)$ is again the local Tolman temperature $T=T_{H}/e^{\phi}$, we have $\nabla T/T = -\nabla \phi(r)$, related to gravitational acceleration.
The force $X_G$ identified as the inverse-temperature gradient coincides directly with the gravitational acceleration field itself, thus providing a direct link between thermal and mechanical degrees of freedom.

%To derive expressions for heat and entropy currents and critically assess the argument that entropic forces arise from extremizing entropy under , 

% Hawking radiation or the second is a dimensionless gravitational-potential gradient normalised by $T$.$\nabla(1/T)$ is dominated by the Hawking temperature gradient near the horizon, 

% Hawking radiation or the second is a dimensionless gravitational-potential gradient normalised by $T$.$\nabla(1/T)$ is dominated by the Hawking temperature gradient near the horizon, 

\paragraph{Onsager Repiciprocity}
In these definitions, the entropy/heat flow $J_S$ and the shell's motion $v$ are coupled because a moving shell can do work and produce heat, and a heat flow can in turn influence pressure/forces via buoyancy or radiation pressure. The Onsager matrix $L_{ij}$ captures this coupling. We can write the linearized equations explicitly as:

\begin{equation}
\begin{pmatrix}
J_S\\[6pt] v
\end{pmatrix}
=\begin{pmatrix}
L_{11} & L_{12}\\[6pt]
L_{21} & L_{22}
\end{pmatrix}
\begin{pmatrix}
X_T \\[6pt] X_G
\end{pmatrix} \,,
\label{Onsager-matrix}
\end{equation}
with $L_{12}=L_{21}$. Here $L_{11}$ is related to the thermal conductivity (governing heat flow due to a temperature gradient), $L_{22}$ characterizes the mechanical dissipation (e.g. how easily the shell moves under a force), and $L_{12}=L_{21}$ are cross-coefficients describing gravity-thermal coupling. The off-diagonal terms mean that a gravitational force can induce an entropy current, and conversely a temperature gradient can induce mechanical work. Physically, this is the key to treating gravity as an entropic force: the tendency of the system to dissipate negative entropy and increase total entropy (via heat flow into the bath) is directly linked to the tendency to move the shell inward (mechanical motion).

 Crucially, the entropy production in these gravitationally influenced near-equilibrium steady states (NESS) is also quadratic in the driving forces near equilibrium, and its positive definiteness ensures compliance with the second law of thermodynamics.

\subsection{Steady-State Gradient and Entropic Force}
We now examine the static steady state where the shell is held fixed by $F_{\rm ex}$ and no net currents flow. In this quasi-equilibrium situation, all fluxes vanish ($J_S = 0$ and $v=0$), while the system maintains small but nonzero intensive gradients. Setting $J_1=J_2=0$ in (\ref{Onsager-matrix}) gives two homogeneous linear equations for the forces $X_T$ and $X_G$. Nontrivial solutions (gradients) exist if and only if the forces are proportional and opposite, ensuring the flux cancellation. In particular, from the second equation $0 = L_{21}X_T + L_{22}X_G$, we find:
\[
X_G = -\,\frac{L_{21}}{L_{22}}\;X_T \,.
\]
Using $L_{21}=L_{12}$, the same relation emerges from the first equation. This proportionality between $X_G$ and $X_T$ is nothing but the condition for detailed balance in the stationary state. Substituting the definitions, we obtain a direct relation between the gravitational force and the temperature gradient. In our case, $X_T = \nabla(1/T)=F_{ext}$ and $X_G \propto F_{\rm grav}$. Thus the no-current condition yields:
\bea
F_{\rm grav} \;\propto\; -\nabla\Big(\frac{1}{T}\Big) \,.
\eea

More rigorously, one can derive the standard Tolman equilibrium condition from the above: eliminating the proportionality constant by combining with the hydrostatic condition, we recover $\nabla T/T = -\,\nabla\phi$ (with $\phi$ the gravitational potential). This is exactly the Tolman-Ehrenfest law for a self-gravitating system in thermal equilibrium. In other words, the reason a Hawking temperature gradient exists in \textcolor{red} {
}
 open gravitational system is that it is required to prevent any net entropy current when the shell is static. The local temperature rises as one falls deeper in the gravitational potential well, such that no spontaneous heat flow occurs. Mathematically, the vanishing of the total entropy production $J_1 + J_2 = 0$ is equivalent to this condition. Onsager reciprocity thus leads us to the {equilibrium state} defined by balanced forces $X_T$ and $X_G$. And quasi-static process means the NESS when moving with small velocity.

Now we turn to the dynamics when the system is allowed to evolve with the external force, then the shell would begin to fall quasi-static (i.e. $v\neq 0$) and entropy would be produced ($J_S\neq0$). Onsager's coupled equations then describe how the incipient motion feeds into entropy increase, and vice versa. Solving (\ref{Onsager-matrix}) for the fluxes in terms of forces, we have in particular the entropy current (to linear order in the perturbations):
\bea
J_S = L_{11}X_T + L_{12}X_G \,.
\eea
We can substitute $X_G$ from the above equilibrium relation to find the net entropy production rate when the detailed-balance condition is slightly violated. Writing $X_G = -\frac{L_{12}}{L_{22}}X_T + \delta X_G$ (where $\delta X_G$ is a small imbalance force not countered by $X_T$), we get
\bea
J_S = \Big(L_{11} - \frac{L_{12}^2}{L_{22}}\Big)X_T + L_{12}\,\delta X_G \,.
\eea
The coefficient in front of $X_T$ is positive by the Onsager-Casimir symmetry (it is essentially a transport coefficient for thermal diffusion in the presence of gravity). 

 {Thus if the shell is released quasi-static ($\delta X_G$ becomes nonzero in the direction of gravity), one finds $J_S > 0$. In physical terms: when the shell falls inward with dissipation, entropy is generated,} flowing from the shell into the black hole or surrounding bath. This entropy production can be identified with the negentropy of the shell being dissipated. The external agent's force $F_{\rm ex}$ had been holding back this tendency; once removed, the negentropy is irreversibly converted into thermal entropy. The second law is upheld: the entropy of the external bath increases (by $\delta S_{\rm bath}>0$), exactly offsetting the decrease of the system's negative entropy $N$. The Onsager framework makes this causal direction clear: the sign of $L_{12}$ (which is positive, since a positive $X_G$ in the direction of gravity combined with a positive $X_T$ gradient yields a positive $J_S$) ensures that entropy flows out of the shell when gravity pulls it in.

\bea
F_g \;=\; T_H \,\frac{d(\Delta S_{\rm BH})}{dx} \;=\; T_H\,\frac{d(S_{\max}-S)}{dx} \;=\; T_H\,\frac{dN}{dx} \,.
\eea
Combining the above results, we can now formulate the entropic force law for gravity in our setup. Consider an infinitesimal quasistatic process where the shell moves inward by $dx$ (a small radial displacement) under the influence of gravity, while an equivalent amount of heat $dQ$ is extracted by the external agent to keep the process slow. The work done by gravity is $dW_g = F_gdx$ (with $F_g$ the gravitational force on the shell), and the heat absorbed by the bath is $dQ$. By energy conservation, $dQ = -dW_g$. Meanwhile, the entropy increase in the bath is $dS_{\rm bath} = dQ/T_H$ (using the Hawking temperature as the characteristic temperature at which heat is dumped into the environment). The negentropy change of the system is related to the change in the black hole's horizon entropy: as the shell moves inward, the horizon area grows by an amount $\Delta A$, corresponding to an entropy increase $dS_{\rm BH} = \frac{\Delta A}{4G}$. If the shell's mass is $m$, one finds $dS_{\rm BH} \approx \frac{m,dx}{T_H}$ (since $m/T_H$ is the entropy the shell would add if fully absorbed 

 This lost ``potential'' entropy of the shell is exactly the negative entropy $dN = -d(S_{\max}-S) = -d(\Delta S_{\rm BH})$. Equating the heat flow to entropy changes, we have $dQ = T_H dS_{\rm bath}$ and $-dQ = T_H d(\Delta S_{\rm BH})$ (the minus because the shell's entropy deficit decreases as it moves in). Therefore $dW_g = T_H d(\Delta S_{\rm BH})$. Dividing by the distance $dx$ inward, we obtain:
\bea
F_g \;=\; T_H \,\frac{d(\Delta S_{\rm BH})}{dx} \;=\; T_H\,\frac{d(S_{\max}-S)}{dx} \;=\; T_H\,\frac{dN}{dx} \,.
\eea
This remarkable relation says that the gravitational force is equal to the Hawking temperature times the spatial gradient of negative entropy.

\section{Negative entropy}
\label{sec:negentropy}

In the previous sections, we introduced the idea that gravity is driven by a gradient of \emph{negative entropy} (or entropy deficit) $N$. We now formalize the definition of $N$ and further elucidate its physical meaning. Negative entropy is defined as the shortfall of a system's entropy relative to the maximum possible entropy it could have for the given macroscopic constraints (energy, volume, charges, etc.). 

In formula form, if $S_{\max}$ is the entropy of the hypothetical equilibrium state with those constraints and $S$ is the system's actual entropy, then we define
\begin{equation}
N \equiv -\Big(S_{\max} - S\Big)\,.
\label{eq:Ndef}
\end{equation}
By construction, $N \le 0$, with $N=0$ if and only if the system is in the maximally entropic state (thermal equilibrium saturating all relevant entropy bounds). A system with $N<0$ contains an ``entropy deficit''-it is more ordered, or contains more information, than an equilibrium system of the same energy. This negentropy $-(S_{\max}-S)$ quantifies the potential work or free energy associated with the system's deviation from equilibrium.

 For a gravitational system, the maximum entropy state for a given energy and boundary conditions is often a black hole. Thus, if a mass-energy distribution is not fully collapsed into a black hole, the total entropy is lower than the hypothetical black hole entropy, and we can identify a negative entropy proportional to the missing horizon area. In particular, because the Bekenstein-Hawking entropy of a black hole is $S_{\text{BH}} = A/(4G)$ (with $A$ the horizon area), a small deviation $\Delta A$ of horizon area from its maximum possible value corresponds to a negative entropy 
\begin{equation}
N \;\approx\; -\,\frac{\Delta A(\mu_a)}{4\,G}~,
\label{eq:N-dA}
\end{equation} 
up to factors of order unity (we take $c=\hbar=k_B=1$ units). Equation \eqref{eq:N-dA} makes clear that a decrease in horizon area (or a horizon area that is smaller than the maximal one for the given energy) corresponds to a negative entropy $N<0$. We will see that this quantity $N$ is closely tied to gravitational binding energy and can be interpreted as an information measure in gravitational thermodynamics.

Between $\rho$ and its maximally coarse-grained state $\bar{\rho}$ is just the negative of the entanglement entropy bound
\bea
N=-\lim_{upper}\Delta S
\eea
and therefore
\bea
N=-S(M,m)
\eea
to reformulate the negentropic formulas.

\subsection{Entropic direction}

When the local Hawking temperature is very low such that the distribution factor $e^{-\omega/T}\ll1$, $\Delta\braket{H_A}$ for the co-moving observer Alice's local measurement $H_A$ stays almost the same when Alice moves between nearby local static trajectories labeled by $\lambda'$ and $\lambda$ in such frequency-fixed process:% with respect to the parameter $\lambda$
\bea
\Delta\braket{H_{A'}}\approx \Delta\braket{H_A}
\eea
so the gravitational redshift effect to the frequency is canceled.
And the entropy bound varies in this the isoenergic process approximates to
\bea
d\Delta S\approx\frac{\Delta\braket {H}}{T_H}d V(r)
\eea
Thus this process approximates to the isoenergic process in low temperature limit, and (\ref{uniexforce}) leads to a minus sign in
\bea
dW_g\approx-T_Hd\Delta S=-dQ\,
\label{eq:dWg}
\eea
which conforms that the opposite direction to the variation of the entropy bound $\Delta S$, while $F_{ex}$ is along the direction of the maximal heat flux $dQ=T_Hd\Delta S$ into the bath during such processes.

What's important is that $F_{ex}$ is actually a thermodynamic force in resistance of $F_g$. Which means it is a totally dissipator to create more entropy. Without such dissipator, we could not tell the direction, as there is neither $F_g$ nor entropic variation in the inertial frame.

We can recast the above result in a more suggestive form. Rearranging \eqref{eq:dWg}, we have $\mathrm{d}Q = T_H\,\mathrm{d}(\Delta S) = -\,\mathrm{d}W_g$. Since $\mathrm{d}Q = T_H\,\mathrm{d}S_{\text{bath}}$ (increase in bath entropy), and $\mathrm{d}W_g = F_g\,\mathrm{d}x$, we get $F_g\,\mathrm{d}x = -\,T_H\,\mathrm{d}(\Delta S)$. Dividing by $\mathrm{d}x$, this is 
\begin{equation}
F_g \;=\; -\,T_H\,\frac{\mathrm{d}(\Delta S)}{\mathrm{d}x}~.
\label{eq:F-entropy-gradient}
\end{equation}
But $-\mathrm{d}(\Delta S)/\mathrm{d}x$ is just $\mathrm{d}N/\mathrm{d}x$, the spatial gradient of the system's entropy (since the Casini-Bekenstein bound is actually $\Delta S = S_{\max}-S$ for fixed $S_{\max}$ in this near-equilibrium setup). Thus \bea
\bold F_g = T_H\,\nabla N\nn,
\eea
an equation that resembles the entropic force formula proposed in analogies of gravity with entropic gradients. However, we emphasize that this equality holds only in the presence of the external agent that allows $\mathrm{d}S$ to be realized as an actual entropy increase. In our derivation, $T_H \nabla N$ really stands for $T_H \nabla (S-S_{\max})$ initially. When the process occurs, $\nabla N$ is converted to $\nabla S_{\text{bath}}$ essentially.

%close and open The fine-grained entropy of a closed system will remain conserved.

 %of covariant entropy bound , {Jacobson:1995ab,Jacobson:2015hqa}

\subsection{AdS/CFT Dual Perspective}

Finally, we consider the meaning of negative entropy from the viewpoint of the AdS/CFT correspondence (holography), which provides a dual description of gravitational physics in terms of a conformal field theory (CFT) living on the boundary. In the AdS/CFT framework, black hole entropy is understood as the entropy of entanglement (or thermodynamic entropy) in the dual field theory. A classic example is the eternal AdS black hole, which is dual to the thermofield double state of two identical CFTs. The von Neumann entropy of one CFT (obtained by tracing out the other) equals the Bekenstein-Hawking entropy of the black hole, given by the area of the horizon (more precisely the area of the extremal surface homologous to the boundary). In such a state, the entropy is maximal for the given energy in the sense that it saturates an analog of the Bekenstein bound: the state is in thermal equilibrium and there is no accessible negentropy.

In the static AdS black hole context, where
\bea
f(r)=1+\frac{r^{2}}{L^{2}}-\frac{16\pi GM}{(d-2)\Omega_{d-2}r^{d-3}},
\eea
 we usually consider equilibrium scenarios. However, we can still imagine a thought experiment in AdS: start with an AdS black hole in equilibrium with its Hawking radiation (which in AdS reflects off the boundary and returns, achieving equilibrium in the Hartle-Hawking state). This is dual to a thermal state in the CFT. Now introduce a perturbation in the CFT that lowers its entropy while keeping energy the same (for instance, couple in an ordered system or do a quench that produces a less-than-thermal state). The dual will be that the bulk now has some structure (like a shell of lower entropy matter) which results in the horizon area being smaller than it would be for a purely thermal state of that energy. The difference in horizon area is accounted for by correlations or excitations outside the horizon. Over time (if the system is left isolated within AdS), one expects that this perturbation will thermalize: the shell will fall in or spread out, eventually increasing the horizon area to the appropriate value and restoring a thermal state. In doing so, the CFT entropy will increase to the thermal value. The negative entropy that was initially present is thus dissipated away in the closed system by mixing between the shell and the Hawking radiation (in AdS, Hawking radiation never escapes, so it effectively provides the thermal bath). In a sense, AdS acts like a confining box that ensures eventual thermalization. In contrast, if we had an open system (like AdS coupled to an external bath or flat space with an environment), the negative entropy can be transferred out rather than simply thermalized in place. For instance, in the eternal AdS black hole with a one-sided perturbation, one can allow an energy flux out through the boundary, analogous to extracting work. This would reduce the mass of the black hole and carry away the information in outgoing radiation to the boundary CFT (which now is not isolated). Holographically, one side of the TFD will lose entropy faster than just by energy loss, reflecting the information carried away.

When we perturb the black hole by adding some matter (for instance, a shell or an excitation) outside the horizon, the dual CFT state is no longer the exact thermofield double; it now carries some additional information or correlation that distinguishes it from a purely thermal state. In the CFT, one way to quantify this is via \emph{relative entropy} or deviations from the entanglement entropy of the vacuum/thermal state. Casini's version of the Bekenstein bound , which we employed in section \ref{Close-Eq}, is in fact a statement about the positivity of relative entropy: $S(\rho||\sigma) = \Delta \langle H_A \rangle/T - \Delta S \ge 0$ for a state $\rho$ (with energy $\Delta \langle H_A \rangle$ above vacuum in region $A$) relative to the vacuum (or a thermal state $\sigma$). Saturation of this inequality ($S(\rho||\sigma)=0$) occurs if and only if $\rho=\sigma$, i.e. the state is exactly the thermal state (no additional information). 

In the gravitational dual, this saturation corresponds to the situation where the space outside the horizon is in the Hartle-Hawking state (no additional excitations) and the entropy is exactly $A/(4G)$. Whenever there is a perturbation ($S(\rho||\sigma)>0$), the inequality $\Delta S \le \Delta \langle H_A \rangle/T$ is strict: $\Delta S < \Delta \langle H_A \rangle/T_H$. Equivalently, the actual entropy is below the bound, $\Delta S_{\text{actual}} = S(\rho) - S(\sigma) < \Delta \langle H_A \rangle/T_H = \Delta S_{\max}$. This difference $\Delta S_{\max} - \Delta S_{\text{fine-grained}}$ is precisely the negative entropy $|N|$. In the CFT, it measures the distinguishability of the state $\rho$ from the thermal state $\sigma$; it indicates that $\rho$ has additional information (for example, correlations or purity) beyond what the thermal state has. In our gravitational context, that additional information is what is stored in the configuration of the shell and gravitational field.

\subsection{Holographic Extremal Surfaces}
The Ryu-Takayanagi (RT) formula \cite{Ryu:2006bv} and its generalizations HRT \cite{Hubeny:2007xt} provide a geometric way to calculate entanglement entropy in the CFT by finding areas of extremal surfaces in the bulk. For a given boundary region (say, one entire AdS boundary in the two-sided black hole scenario), the RT surface in the unperturbed thermofield double is the event horizon (more precisely the bifurcate horizon or the area at $r_H=2GM$ slice). When we perturb the system (e.g. put a shell outside one side of the black hole), the extremal surface that computes the entropy of the boundary region will generally shift or deform. If the shell carries information (i.e. it is not just thermal noise but a specific macrostate), the entanglement entropy of the boundary region will \emph{decrease} compared to the unperturbed case, because the overall state is less mixed (the two sides plus shell are not in perfect thermal equilibrium). In the bulk, one may say that the extremal surface dips inside the horizon or moves in such a way that its area is smaller than the area of the unperturbed horizon. This reduction in area corresponds to the negative entropy from the bulk point of view, which is the ``entropy deficit'' between apparent horizon at $r_{a}$ and event horizon at $r_{h}$
\bea
-N=\Delta S=\Delta A/4G=\frac{\Omega_{d-2}}{4G}(r_{a}^{\,d-2}-r_{h}^{\,d-2}).
\eea

According to the Ryu-Takayanagi prescription, the entropy of a boundary region \(A\) equals \(\mathrm{Area}(\gamma_A)/(4G)\), where \(\gamma_A\) is the bulk minimal surface homologous to \(A\).  For the entire boundary in the eternal black hole geometry, \(\gamma_A\) coincides with the bifurcation surface of the horizon.  Adding the shell perturbs the metric and pushes \(\gamma_A\) outward; its area increases precisely by the same \(\Delta A\) found above.  Thus the holographic entropy shift \(\Delta A/(4G)\) matches the Bekenstein-Hawking result, confirming that injecting energy \(m\) raises the CFT entropy by \(m/T_H\). 

In fact, one can make this more precise by using the concept of a \emph{quantum extremal surface} (QES), which includes not just the area term but also the entropy of quantum fields outside that surface. The condition for a QES extremizing 
\bea
S_{\text{gen}} = \frac{\text{Area}}{4G} + S_{\text{out}}
\eea
 encapsulates the balance between including more horizon area versus excluding more outside entropy. A perturbation like a shell represents a situation where the naive horizon is no longer the minimal QES; instead, a surface outside the horizon (encompassing the shell) might become relevant. The difference in area between the horizon and the new extremal surface, in combination with the entropy of fields, gives a measure of $N$. In simple terms, the presence of the shell means that the entanglement entropy of the Hawking radiation or fields outside is reduced (since some degrees of freedom are in a pure or correlated state with the shell), and thus the total generalized entropy is lower than it would be if everything were maximally mixed. This is directly the holographic dual of having $N<0$.

To summarize the AdS/CFT perspective: negative entropy in the bulk corresponds to the statement that the boundary state is not maximally entropic (has correlations/information relative to a thermal state). The gradient of negative entropy, which we associated with the gravitational force, has a dual interpretation as well: a spatial variation of $N$ in the bulk corresponds to a spatial (or scale) variation of entanglement in the CFT. For example, the fact that $F_g \propto \nabla N$ in bulk translates to something like an entropic force in the CFT, though the precise dual statement might be more subtle, involving how the modular Hamiltonian gradients relate to physical forces on subsystems . Our usage of Onsager relations and local temperature gradients in Section \ref{Close-Eq} can also be given a CFT interpretation: a temperature gradient in the bulk corresponds to a gradient in the one-point functions or densities in the CFT, which via Einstein's equations (holographically) is linked to the stress-energy gradient and hence acceleration (force). Through the duality, the thermodynamic force $F_{\mu} = T \nabla_{\mu}(K_0 - K_1)$ ) that we derived from entanglement first law in Section \ref{Close-Eq} is a direct translation of the gravitational entropic-force law into field theory terms. There, $K_0$ was the modular Hamiltonian of the unperturbed (maximally entropic) state and $K_1$ that of the perturbed state; $K_0-K_1$ encodes the ``negentropy'' of the perturbation. The emergence of $F_{\mu}$ from the difference $K_0-K_1$ is essentially the CFT seeing the gradient of negative entropy as a force.

\section{Discussion}
%In this work we focused on near-equilibrium steady states (with linear response via Onsager relations), since 

We have presented a new framework for emergent gravity that centers on negative entropy (negentropy) as the driver of spacetime curvature, distinguishing it from conventional entropic gravity theories. In this model, a gravitating system out of complete equilibrium-exemplified by a black hole with an external thin shell-was analyzed using both Euclidean path integrals and non-equilibrium thermodynamics. Several key results were achieved. First, we derived an explicit expression for the entropy change due to a shell perturbation and showed how this relates to a negentropy flux. Specifically, when a shell of mass $m$ is held at radius outside a black hole of mass $M$, the horizon's area and entropy respond by a fractional amount ($\Delta A$ and $\Delta S$), and we identified this $\Delta S$ with $S(M,m) = m/T_H(M)$. This quantity represents the entropy that has not yet been realized by the black hole-effectively, the entropy ``missing'' because the shell's energy remains organized. We demonstrated that the flux of negative entropy from the shell to the horizon (as the system equilibrates) is what generates the entropic force of gravity. In practical terms, the gravitational attraction can be understood as the system's tendency to increase its entropy by assimilating the shell's negentropy. Our use of Onsager's reciprocity made this precise: we showed that a steady entropy current $dS/dt$ is driven by the Hawking temperature gradient, and the corresponding thermodynamic force (the external force needed to hold the shell, or equivalently the gravitational pull) satisfies a linear response relation. The result is an entropic-force law with a firm basis in thermodynamics, rather than an assumed entropy extremum. We also found an integrable structure in the geometric description-by treating the thin shell in incremental layers, we could integrate the contributions of each layer's entropy change and recover the full Bekenstein--Hawking entropy of a black hole. This integrability implies consistency: the negentropy contributions from multiple shells add up exactly to the expected limit (for instance, stacking many infinitesimal shells until the mass $m$ is fully absorbed reproduces the entropy of a black hole of mass $M+m$).

%The beginning with the entanglement 1st law itself, has the power to rescue gravitational force from being totally chaotic and entropic, where $W_g$ is emergent naturally as the source supplying negentropy while $dQ=TdS$  transforming the gravitational potential, with an Hawking temperature gradient as input.

A central theme of our conclusions is the dual role of the negative entropy $N$ unveiled by this work. We have shown that $N$ can be viewed both as a geometric quantity and as a coarse-graining measure of information. Geometrically, $N$ is proportional to the reduction in horizon area (or the area of an extremal surface) due to the presence of an ordered matter configuration like the shell. It is literally encoded in the Euclidean geometry of the solution -- for instance, through the deficit angle or the on-shell action difference caused by inserting the shell. At the same time, $N$ captures the degree of organization or correlation in the system: it measures how much smaller the entropy is compared to a reference state of maximal entropy. In our example, the shell's existence means some degrees of freedom are in pure states or entangled with the shell, rather than contributing to thermal entropy. This lack of entropy is $N$ -- a positive negentropy indicating information content. Thus, $N$ serves as a bridge between microscopic information and macroscopic geometry. We emphasized that this aligns well with the spirit of holography: the area of a surface (a geometric concept) corresponds to entropy of degrees of freedom (an information concept). Our work makes this correspondence concrete for a dynamic, near-equilibrium setting. In summary, the negative entropy $N$ introduced here is not an abstract quantity; it is built into the spacetime's structure and simultaneously reflects the system's entanglement pattern. This dual interpretation strengthens the connection between gravity and quantum information, suggesting that spacetime curvature in our framework is literally an information-driven phenomenon.

The gravitational attraction is also a reflection of entanglement structure of spacetime. Our approach started with the entanglement first law itself, which already has the power to rescue gravitational force from being viewed as purely chaotic or thermally emergent. Instead, we found that the gravitational potential (work term $W_g$) naturally arises as the source supplying \emph{negentropy}, while the equality $dQ = T\,dS$ is transformed into the gradient of gravitational potential with a Hawking temperature input. We may extend the entanglement structure of the matter in gravitational field in the future study.

Since many physical scenarios - such as gravitational collapse, Hawking radiation back-reaction, or cosmological expansion - are genuinely out of equilibrium. It would be fascinating to investigate how entropic forces and negentropy play out in these regimes, possibly requiring nonlinear generalizations of Onsager's theory or invoking fluctuation-dissipation theorems in a gravitational setting. The notion of an ``entropy deficit'' driving dynamics might also shed light on processes like information retrieval from black holes (where negentropy is returned to the exterior via Hawking radiation). Additionally, our approach could be applied to open quantum systems and analog gravity setups: one could imagine laboratory systems where an entropy gradient mimics gravity, allowing experimental probes of negentropic forces. Overall, by reframing gravity as an emergent phenomenon sustained by negative entropy and irreversible thermodynamics, we have provided a fresh perspective that integrates geometry, thermodynamics, and quantum information. The encouraging results here - from the successful matching of Euclidean calculations with holographic entropy, to the consistency with gravitational thermodynamic laws - suggest that negentropic gravity is a fruitful paradigm. We anticipate that further research along these lines, including holographic RG analyses and non-equilibrium field theory explorations, will deepen our understanding of how spacetime and gravity emerge from the fundamental interplay of entropy, information, and quantum statistical mechanics.
\acknowledgments

We would like to thank Peng Cheng, Shanming Ruan, Jia Tian, Yixu Wang, Cheng Peng, Zhenbin Yang for useful discussion. This research was first presented on the Quantum Convergence: Information and Gravity 2024 in Mogan mountain hold by IASTU of Tsinghua University.
Y.A was grateful for Yu-sen An, Baofei Li, Jing Liu, Yuanzhu Wang, Chao Zhang for support and help. 
\appendix
\section{The Emergence of inertial force}
\label{EmIF}
Based on the entanglement first law, we could take the parameters such as temperature $T$ in $K_0=H_0/T$ out of the derivative
\bea
W_g+TdS_1=d\braket H_0 \,.
\eea
 during certain processes.
 
We can compare the first law 
\bea
T d S_0=d \braket{H_A}_0\,,\\
d W_g+Td S_1=d \braket{H_A}_1\,.
\label{heat}
\eea
 and the work term can be simplified to from temperature gradient
\bea                                                                                                                                                                                                                                                                                                                                                                                                                                                                                                                                                                                                                                                             
dW_g=-T\times {\Delta\braket{H_A}}d\frac{1}{T}\,,
\label{dW-}
\eea
which doesn't depends on the detail form of the operator $O$.
 the local temperature field %$\Delta \braket{H} \rightarrow m$ and 
$T=\frac{T_0}{V}$ for this temperature-changing process leads to the inertial force
\bea
\textbf{F}_g=-T\times \frac{\Delta\braket{H_A}}{T_0}\nabla_\mu V\,.
\label{dF-}
\eea
This  formula is exactly opposite to the external force formula (\ref{entforce}) with the entropic gradient (\ref{entgrad}).
Noticing the minus sigh in (\ref{dW-}) and (\ref{dF-}), the approach using the entanglement first law will reproduce the inertial force. Following, we show the entropic force formula (\ref{entforce}) together with the entropic gradient will reproduce the external force, for  Newton's 2nd law and gravitational force, as we expect.

\subsection{Emergence of Newton's 2nd Law in Rindler Space}

In the coordinate $\{\eta,\xi\}$, the metric of Rindler space is
\bea
ds^2=e^{2a\xi}(-d{\eta^2}+d{\xi}^2)\,,
\label{RindlerI}
\eea
for the right Rindler wedge of the Minkowski spacetime.

Every orbit $\xi\equiv const$ corresponds to one of the different accelerating observers following a boost killing vector $\d_\eta$. Those accelerating orbits share the same Rindler horizon $H^\pm$ as well as the same causal development, which is the right Rindler wedge.

The redshift factor is
\bea
V(\xi)=\sqrt{-\chi_\mu \chi^\mu}=e^{a\xi}
\eea
where $\chi_\mu$ is the killing vector.

The surface gravity of the Killing horizon of the wedge is just
$\kappa=a$, so the Unruh temperature \cite{Unruh:1976db} is
\bea
T=T_U=\frac{a}{2\pi}\,,
\eea
where the parameter $a$ is also the acceleration of the observer following the orbit $\xi\equiv0$. 
%According to the Unruh effect\cite{Unruh:1976db}, 

From the proposed entropic gradient expression (\ref{entgrad}), we will get
\bea
\nabla_\mu S=
\frac{m}{T_U}\delta^{\xi}_{\mu}\d_\xi V(\xi)
=
\delta_{\mu}^{\xi}2\pi m e^{a\xi}\,,
\eea
and the entropic force formula (\ref{entforce}) produces
\bea
F_\mu=T_U\nabla_\mu S=\delta_{\mu}^{\xi} m a
 e^{a\xi}\,,
\eea
where the covariant $\delta_{\mu}^{\xi}$ shows the force is in the direction to switch the orbit towards the one with higher acceleration. 
So the external force $F=\sqrt{F_\mu F^\mu}$ is
\bea
F=ma\,,
\eea
which exactly agrees with Newton's 2nd Law.

\subsection{Emergence of Gravitational Force}

We set a stationary background of asymptotic flat Schwarzschild black hole with the metric
\bea
ds^2=-(1-\frac{2GM}{r})dt^2+{1\over(1-\frac{2GM}{r})}dr^2+{r^2}d\Omega^{2}\,,
\eea
in the global coordinate. We ignore the back-reaction from our test particle to the geometry.

The redshift factor is
\bea
V(r)=\sqrt{-\chi^\mu\chi_\mu}=\sqrt{-g_{00}}=\sqrt{1-\frac{2GM}{r}}\,,
\eea
the entropic gradient is
\bea
\nabla_\mu S &=&\frac{1}{T_H}\frac{GM}{r^2\sqrt{1-\frac{2GM}{r}}}\delta^r_\mu\,,
\label{Sgenr}
\eea
and the local measure the Hawking temperature for the static observer with $r\equiv const$ is
\bea
T=\frac{T_H}{V(r)}\,.
\eea
So the entropic force formula reproduces
\bea
F_\mu=T\nabla_\mu S=\frac{GMm}{r^2(1-\frac{2GM}{r})}\delta^r_\mu\,.
\eea
For the observer at infinity, the force amounts $F=V(r)\sqrt{F_\mu F^\mu}=\frac{GMm}{r^2}$.

Notice that it is directly covariant results calculated in General Relativity, see textbook \cite{Carroll:2004,Wald:1984}. These results agree with the local external force $\textbf{F}_{ext}$ calculated in General Relativity
\bea
\textbf{F}_{ex}=ma_\mu
\eea
with $a^\mu=U^\nu\nabla_\nu U^\mu$, for the static observer whose four-velocity $U^\mu$ is proportional to the time-translation Killing vector $\d_t$. %$a_\mu=\frac{GM}{r^2(1-\frac{2GM}{r})}\delta^r_\mu$,
 %Obviously, the $inertial force $\textbf{F}_g=-F_\mu$ approximates Newton's Law of Gravity in the same way as Einstein Gravity. 
\paragraph{Near-Horizon Limit is not generalizable}
We note that, to form a general entropic mechanism, the local Hawking temperature $T(r)=\frac{T_H}{V(r)}$ plays the ordinary role of the Unruh temperature $T_U$ in Entropic Gravity theories.  And our results are directly consistent with the gravitational force, not just in the near-horizon region. 

In the near-horizon limit, the black hole geometry approximates the Rindler space while the local Hawking temperature approximates the Unruh Temperature, that's why an entropic mechanism works directly in generic situations can be applied to the near-horizon region, not the other way around.

\section{Euclidean Path Integral}
The full (renormalized) Euclidean action of the system includes the gravitational action in each region plus the shell's own action.  In the semiclassical (saddle-point) approximation one has
\bea
I_E = I_{\rm grav} + I_{\rm matter}\,,
\eea
where
\bea
I_{\rm grav}=-\frac{1}{16\pi G}\int_{M}R\sqrt{g}d^4x + \frac{1}{8\pi G}\int_{\partial M}(K-K_0)\sqrt{h}d^3x
\eea
and $I_{\rm matter}$ is the Euclidean matter action.  Here $K_0$ is the extrinsic curvature of a flat-space reference (Minkowski) boundary, whose inclusion implements the background subtraction.  Since both inside and outside regions are vacuum Schwarzschild, their Ricci scalars vanish and only the boundary terms contribute.  
\subsection{Extrinsic curvature on the shell}
Since each region is a vacuum solution ($R_{\mu\nu}=0$), the bulk Einstein-Hilbert term vanishes on shell.  All contributions to the on-shell Euclidean action come from Gibbons-Hawking-York (GHY) boundary terms \cite{Gibbons:1976ue}. 

We thus compute the extrinsic curvature of the spherical shell hypersurface $r=R$ in each region.  A unit normal to the shell (pointing toward increasing $r$) is $n^\mu=\bigl(0,\sqrt{f(r)},0,0\bigr)$, so the trace of the extrinsic curvature is
\bea
K = \nabla_\mu n^\mu \;=\; \frac{1}{2\sqrt{f}}\frac{df}{dr} + \frac{2\sqrt{f}}{r}\,,
\eea
which for metric function $f(r)$ on a sphere of radius $r=R$ gives
\bea
K_\pm(R) \;=\; \frac{1}{2\sqrt{f_\pm(R)}}\,f_\pm'(R)\;+\;\frac{2\sqrt{f_\pm(R)}}{R}\,.
\eea

\bibliographystyle{JHEP} 
\bibliography{EmergedGrav}

\end{document}